\documentclass{revtex4}
\usepackage{graphicx,epsf,epsfig,amssymb,amsmath}
\begin{document}

\title{Equilibrium configurations of fluids and their stability in higher dimensions}

\author{Vitor Cardoso} \email{vcardoso@phy.olemiss.edu}
\affiliation{Department of Physics and Astronomy, The University of
Mississippi, University, MS 38677-1848, USA \footnote{Also at Centro
de F\'{\i}sica Computacional, Universidade de Coimbra, P-3004-516
Coimbra, Portugal}}

\author{Leonardo Gualtieri} \email{gualtieri@roma1.infn.it}
\affiliation{ Dipartimento di Fisica Universit\`a di Roma ``La
Sapienza'' and Sezione INFN Roma1, Piazzale Aldo Moro 2, 00185 Rome,
Italy}

\date{\today}

\begin{abstract}
We study equilibrium shapes, stability and possible bifurcation diagrams of fluids in higher dimensions, held
together by either surface tension or self-gravity. We consider the equilibrium shape and stability problem of
self-gravitating spheroids, establishing the formalism to generalize the MacLaurin sequence to higher
dimensions. We show that such simple models, of interest on their own, also provide accurate descriptions of
their general relativistic relatives with event horizons. The examples worked out here hint at some
model-independent dynamics, and thus at some universality: smooth objects seem always to be well described by
both ``replicas'' (either self-gravity or surface tension). As an example, we exhibit an instability afflicting
self-gravitating (Newtonian) fluid cylinders. This instability is the exact analogue, within Newtonian gravity,
of the Gregory-Laflamme instability in general relativity. Another example considered is a self-gravitating
Newtonian torus made of a homogeneous incompressible fluid. We recover the features of the black ring in
general relativity.
\end{abstract}

\maketitle

\section{Introduction}
\begin{figure}[ht]
\centerline{\includegraphics[width=10 cm,height=8 cm] {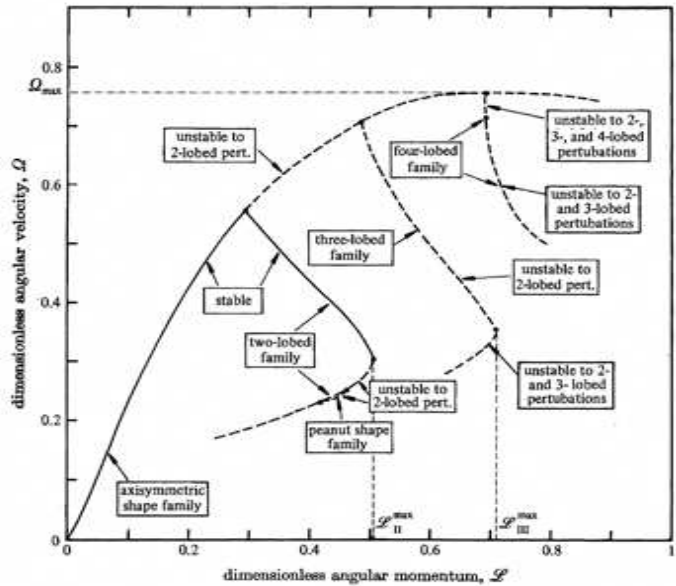}} \caption{The bifurcation and stability
diagram for rotating liquid drops with surface tension (self-gravity is neglected). Shapes are stable along
solid branches, unstable along broken ones. The plot depicts dimensionless angular velocity as function of
dimensionless angular momentum. The dimensionless angular velocity is $\Omega=\sqrt{\frac{\rho
\hat{\Omega}\hat{R}}{8T}}$, where all hatted quantities are dimensionfull, $\hat{R}$ is the radius of a sphere
with the same volume as that of the drop, $\rho$ and $T$ are the density and surface tension of the liquid.
Likewise, the dimensionless angular momentum J is defined as $J=\hat{J}/4T \hat{R}^2\left (8T\hat{R}^2
\rho\right )$. Although not shown in the plot, the axisymmetric curve bends back to lower angular velocities
after it has passed the four-lobed family neutral point. Note also that the maximum allowed angular momentum
for {\it stable} shapes is denoted by $J_{II}$ in the diagram. Taken from \cite{brownscriven}.}
\label{fig:brown}
\end{figure}

The shape and stability of liquid drops has been the subject of intense study for more than 200 years, with
exciting revivals occurring more or less periodically. In the early days of the topic, the hope was that
centimeter-sized drops held together by surface tension could be good models for giant liquid masses held
together by self-gravitation. Liquid drops, which could be controlled in the laboratory, should model
self-gravitating stars where gravity would replace surface tension. A great deal of what is now known about
(rotating) liquid drops is due to the work of J. Plateau \cite{plateau} who, despite suffering from blindness
throughout a good part of his scientific life, conducted many rigorous experimental investigations on the
subject (we refer the reader to \cite{plateau} for details). Starting with a non-rotating drop of spherical
shape, Plateau and collaborators progressively increased the rate of rotation (angular velocity, assuming the
drop rotates approximately as a rigid body). Plateau found that as the rotation rate increased, the drop
progressed through a sequence of shapes which evolved from axisymmetric for slow rotation, to ellipsoidal and
two-lobed and finally toroidal at very large rotation. These experimental works have been repeated with
basically the same outcome, even in gravity-free environments \cite{rotatingdropexperiments} (see also
\cite{rotatingdropexperiments2}). Theoretical analysis of Plateau's experiments were started by Poincar\'e
\cite{poincaresurfacetension}, who proved that there are two very different possible solutions to the
equilibrium problem: one is axisymmetric like disks or torii and the other is comprised of asymmetric shapes
with two lobes. The analysis was completed by Chandrasekhar \cite{chandra}, improved upon by Brown and Scriven
\cite{brownscriven} (see also \cite{brownscrivenprl,heine}) and is summarized in Figure \ref{fig:brown}, taken
from \cite{brownscriven}.
The Figure shows the (dimensionless, as defined by Brown and Scriven, see caption in Fig. \ref{fig:brown})
angular velocity as a function of the angular momentum of the drop, and it displays the stability and
bifurcation points of the evolution. Let's focus first on the evolution of axisymmetric figures. Starting from
zero angular momentum, where the shape is a spherical surface, one increases the angular momentum. The
resulting equilibrium shape evolves from a sphere through oblate shapes to biconcave shapes (see also
\cite{chandra}). There are no (axisymmetric) equilibrium figures with $\Omega>\Omega_{\rm max}\sim0.753$. The
axisymmetric family does not terminate at $\Omega=\Omega_{\rm max}$, instead it bends back to lower angular
velocity, and there are now {\it two} axisymmetric figures of equilibrium. The steepest descending branches are
torii \cite{rotatingdropexperiments2,gulliver}.

\begin{figure}[ht]
\centerline{\includegraphics[width=8 cm,height=9 cm] {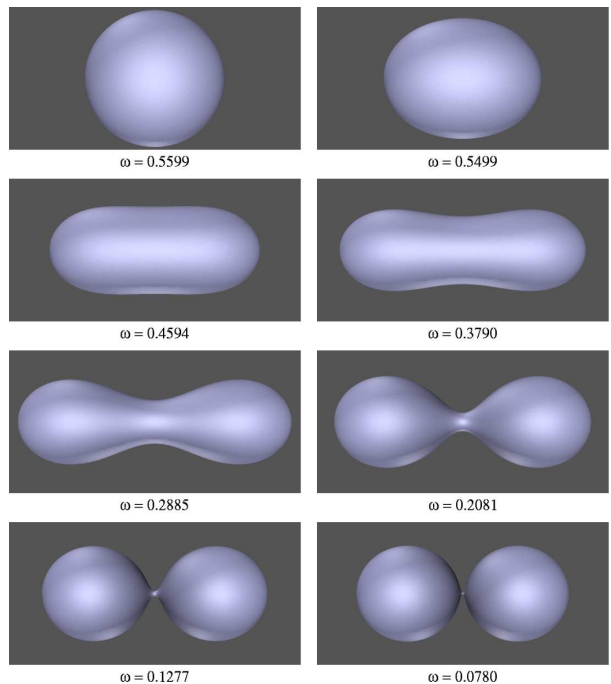}} \caption{This plot shows a sequence of members
of the two-lobed family of drop shapes for several values of angular velocity. The two-lobed drops seem to
approach a limiting surface consisting of two spheres touching each other in one point as $\Omega$ approaches
0. The first shape, with $\Omega=0.5599$, corresponds to the first (two-lobed) bifurcation point in Fig.
\ref{fig:brown}. For $\Omega=0.2885$ for instance, the figure is peanut shaped, and we can situate the
corresponding family in Fig. \ref{fig:brown}. Taken from \cite{phdheine}.} \label{fig:twolobes}
\end{figure}

\begin{figure}[ht]
\centerline{\includegraphics[width=8 cm,height=6 cm] {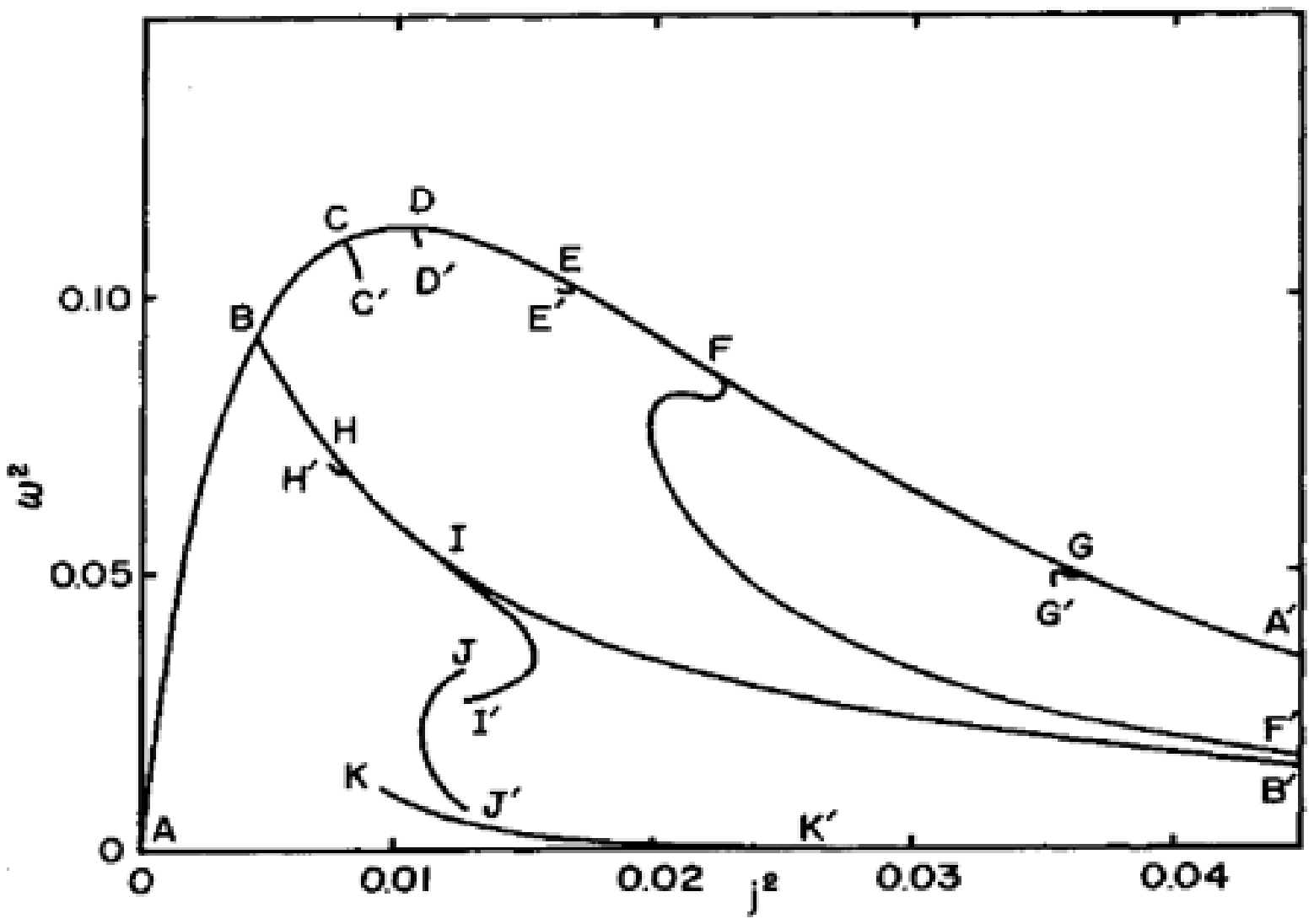}} \caption{The bifurcation diagram for rotating
self-gravitating, incompressible homogeneous stars. Here we show the angular velocity as function of the
angular momentum for several possible families. The data and notation are those of Eriguchi and Hachisu
\cite{eriguchi}. The line $A-A'$ is the MacLaurin sequence, $B-B'$ the Jacobi sequence, $C-C'$ triangle
sequence, $D-D'$ the square sequence, $E-E'$ ammonite sequence, $F-F'$ the one-ring sequence, $G-G'$ the
two-ring sequence, $H-H'$ the pear-shaped sequence, $I-I'$ the dumb-bell sequence, $J-J'$ the Darwin sequence
and finally $K-K'$ is the two point-mass sequence. Note the similarities with Fig. \ref{fig:brown}. See
\cite{eriguchi} for further details. } \label{fig:bifgrav}
\end{figure}

Stability calculations show that all shapes of the axisymmetric family are stable against axisymmetric
perturbations, however a real drop is also perturbed by non-axisymmetric modes. The stability calculations show
that the drop becomes unstable at $\Omega =\Omega_{II}\sim 0.5599$ \cite{brownscriven}, where the axisymmetric
family is neutrally stable to a two-lobed perturbation. This point marks the bifurcation of a two-lobed family
(therefore a non-axisymmetric family). All axisymmetric shapes with $\Omega>\Omega_{II}$ are unstable against
two-lobed perturbations, but there are additional points of neutral stability, which are also bifurcation
points: the three- and four-lobed perturbations are shown in the diagram \ref{fig:brown} but there are others
\cite{heine}. The first, most important non-axisymmetric surface, the two-lobed family, branches off at
$\Omega=\Omega_{II}$, and is initially (near the bifurcation point) stable. In Figure \ref{fig:twolobes} we
show some typical members of that sequence (see \cite{phdheine}, where this figure was taken from, for further
details).
It turns out that the evolution of self-gravitating fluids, i.e., stars, follows a very similar diagram, as
shown in Figure \ref{fig:bifgrav}. Finding the equilibrium configurations of self-gravitating rotating
homogeneous fluids is an extremely important problem, which has met intense activity ever since the first
discussions by Newton and others (see \cite{history} and also the excellent and concise introduction in
\cite{chandrabookellipsoidal}), which was revisited some 50 years ago by Chandrasekhar and co-workers
\cite{chandrabookellipsoidal}. In three spatial dimensions, the most basic figures of equilibrium of rotating
bodies are the MacLaurin and Jacobian sequences \footnote{We consider only Minkowski spacetimes. In the
presence of cosmological constant some new features arise. For instance, the cosmological constant allows
triaxial configurations of equilibrium rotating about the minor axis as solutions of the virial equations
\cite{ellipsoidsdesitter}}. The MacLaurin sequence is a sequence of oblate spheroids along which the
eccentricity of meridional sections increases from zero to one. Jacobi showed that for each value of
$\Omega^2$, the square of the angular velocity of rotation, there are actually two permissible figures of
equilibrium. The other is a tri-axial ellipsoid, belonging to the Jacobi sequence, which branches off from the
MacLaurin sequence. This is the first of several points of bifurcation which distinguish the permissible
sequences of figures of equilibrium of uniformly rotating homogeneous masses. If one starts from a non-rotating
homogeneous fluid ellipsoid in the MacLaurin sequence and increases the angular momentum of the body, one will
first come across the Jacobi branch, where the ellipsoid becomes unstable with regard to the first
non-axisymmetric perturbation \cite{chandrabookellipsoidal} (which are the self-gravitating analogues of the
shapes in Figure \ref{fig:twolobes}). Continuing, still along the MacLaurin sequence and increasing angular
momentum, one next comes to the bifurcation point of other non-axisymmetric figures: the ``triangle'', the
``square'' and ``ammonite'' sequences \cite{chandrabookellipsoidal,eriguchi}. One finally encounters the first
axisymmetric sequence. The bodies of this sequence pinch together at the centre and eventually form Dyson's
anchor-rings, or torus-like configuration \cite{dyson,bardeen,wong,eriguchisugimoto,ansorg}. It is apparent
that the bifurcation diagram for self-gravitating fluids, Figure \ref{fig:bifgrav} is very similar to the one
for non-gravitating fluids with surface tension, shown in Figure \ref{fig:brown}. Thus, whenever the treatment
of self-gravitation proves too difficult, the surface tension model is a good guide.
A realistic star is neither homogeneous nor incompressible and it certainly isn't held together by surface
tension. However, it seems logical to build our understanding using simpler models. We have gone a long way
with simpler related models: another well-known example, where the fluid drop model proved extremely useful, is
Bohr and Wheeler's \cite{bohrwheeler} proposal to describe nuclear fission as the rupture of a rotating charged
liquid drop, where now the surface tension plays the role of nuclear forces. A review with a detailed and
unified discussion of charged nuclei, astronomical bodies and fluid drops can be found in \cite{swiatecki} (a
concise and complete account of the history of this subject, along with some good quality figures can be also
found in \cite{phdheine}). Liquid drops can model self-gravitating systems. Can they model extremely compact,
highly relativistic objects like black holes? The answer, quite surprisingly, seems to be yes
\cite{cardosodiasglrp}. Regarding the event horizon as a kind of fluid membrane is a position adopted in the
past \cite{membraneparadigm}. A simple way to see that this is a natural viewpoint, is to take the first law of
black hole mechanics \cite{BardCartHawk} which describes how a black hole (we will take for simplicity
uncharged, static objects), characterized by a mass $M$, horizon area $A$, and temperature $T=1/(32 \pi M)$,
evolves when we throw an infinitesimal amount of matter into it:
\begin{equation}
dM=TdA \,.\label{firstlaw}
\end{equation}
We now know that the correct interpretation of this law is that black holes radiate, and therefore can be
assigned a temperature \cite{bekenstein,hawking}. The second law of black hole mechanics $dA\geq0$ is then just
the second law of thermodynamics. But we can also argue that Equation (\ref{firstlaw}) can be looked at as a
law for fluids, with $T$ being an effective surface tension \cite{booktension}. The first works in black hole
mechanics actually considered $T$ as a surface tension (see the work by Smarr \cite{smarr} and references
therein), which is rather intuitive: the potential energy for fluids, associated with the storage of energy at
the surface, is indeed proportional to the area. Later, Thorne and co-workers \cite{membraneparadigm} developed
the ``membrane paradigm'' for black holes, envisioning the event horizon as a kind of membrane with
well-defined mechanical, electrical and magnetic properties. Not only is this a simple picture of a black hole,
it is also useful for calculations and understanding what black holes are really like. This analogy was put on
a firmer ground by Parikh and Wilczek \cite{wilczek}, who provided an action formulation of this membrane
picture. There are other instances where a membrane behavior seems evident: Eardley and Giddings
\cite{eardley}, studying high-energy black hole collisions found a soap bubble-like law for the process, while
many modern interpretations of black hole entropy and gravity ``freeze'' the degrees of freedom in a lower
dimensional space, in what is known as holography \cite{holo}. In \cite{kovtun1} it was shown that a
``membrane'' approach works surprisingly well, yielding precisely the same results as the AdS/CFT
correspondence. There have also been attempts to work the other way around: computing liquid surface tension
from the (analog) black hole entropy \cite{surfacetbh}. As a last example of a successful use of the membrane
picture, Cardoso and Dias \cite{cardosodiasglrp} have recently been able to mimic many of the aspects of a
gravitational instability by using a simple fluid analogy. We will comment on this work later on.

Having said this, it strikes one as very odd that the study of simple fluid models has not been extended to
higher dimensions. Extra dimensions, besides the four usual ones, seem to be needed for consistency in all or
many of candidate theories of unification. We have in mind not only string theory \cite{zwie} where the extra
dimensions are compactified to a very small scale, but also recent theories with large \cite{add} or warped
extra dimensions \cite{rs}. Therefore, an understanding of higher dimensional gravitating objects is
imperative. However, even though there are several known exact solutions to Einstein equations in more than
four dimensions describing black objects (with event horizons), nothing or very little is known about the
simpler Newtonian fluid equilibrium configurations, with or without self-gravity.

For instance, higher dimensional black hole solutions with spherical topology have long been known
\cite{tangherlini,myersperry}, which generalize the four-dimensional Kerr-solution to higher dimensions. In
four spacetime dimensions one can show that the topology of a black hole must be a sphere, but that's no longer
the case in higher dimensions. A simple example is to consider extra flat dimensions, for instance ${\rm
Schwarzschild} \times R$, which is still a solution to the vacuum Einstein equations. These extended solutions
can then have different topologies, like a cylinder, and are called back strings \cite{horowitzstrominger}. Yet
another example is the five-dimensional black ring \cite{Emparan:2001wn,Elvang:2003mj}, having the topology of
a torus.

It is curious to note that we are in possession of several exact
solutions to the full Einstein equations and yet nothing is known
about the simpler case of self-gravitating Newtonian fluids or even
fluids without gravity but with surface tension. To be specific, we
found a solution as ``strange'' as a black ring in five dimensions,
yet we do not know how a rotating, self-gravitating Newtonian fluid
behaves as a function of the rotation rate, for higher dimensional
spacetimes. These simpler objects should be much easier to study,
while carrying precious information about their more compact cousins
with event horizons. There are several topics that could and should be
addressed within this research line:

(i) Do the fluid equations admit toroidal-like configurations, which may collapse to black rings for very high
densities? Are these stable?

(ii) Ultra-spinning black holes in spacetime dimensions larger than
four are conjectured to be unstable
\cite{emparanmyers,cardosodiasglrp} for large enough rotation
rates. If the four-dimensional results for self-gravitating fluids or
for fluids with surface tension carry over, this is to be expected. As
we saw in the evolution diagrams, axisymmetric configurations are
unstable for rotation rates larger than a given threshold, decaying to
a two-lobe, asymmetric configuration. This suggests that the end-state
of ultra-spinning black holes is a two lobe, asymmetric state like
those in Figure \ref{fig:twolobes}. This possibility was already put
forward in \cite{emparanmyers}, but the arguments presented here make
this a much stronger possibility. These shapes would then presumably
spin-down due to radiation emission. Can we use simple fluids with
surface tension or with self-gravity to model this process? Can we
estimate the rotation rate at which instability occurs?

(iii) The black strings and p-branes are unstable against a gravitational sector of perturbations. Can we
simulate this instability with fluids? A first step was taken in \cite{cardosodiasglrp}, where it was shown
that a simple fluid with surface tension can account for most of the properties of this instability. Can we add
self-gravity to the model? Can we include rotational and charge effects?

(iv) Can we build the evolution and bifurcation diagrams for fluids with surface tension (or with self-gravity)
in higher dimensions? Can we deduce other solutions to the general relativistic equations using these simpler
models? Can we infer other properties of already known solutions?

These are just some of the topics that could be addressed with simple Newtonian fluid models, but the list
could easily be made much longer. The goal of the present work is to pave the way to some of these possible
developments, but especially to show that these simple models do yield relevant information. We will attack the
problem in its two versions: fluids held together by surface tension and by self-gravity.

\section*{Plan of the paper}
This paper is divided in two parts, Part I in which we deal with fluids held by surface tension and Part II
where we deal with fluids held together by self-gravity.

We start with Section \ref{sec:symm}, where we review the meaning of angular momentum in higher dimensions and
also the symmetries of a rotating body. We also specify the two most important cases (case I and II) we shall
deal with throughout the paper. In Section \ref{sec:hydroeqs} we lay down the basic equations of hydrostatics
and hydrodynamics in higher dimensions and in Section \ref{sec:perturbations} we write the formalism necessary
to analyse small deviations for equilibrium. The formalism just laid down is then used in the next sections. In
Section \ref{sec:eqshapesurface} we study the equilibrium figure of rotating drops. Oscillation modes of
non-rotating spherical drops are derived in Appendix \ref{app:surfacetensiondrop} and constitutes a
generalization to arbitrary number of dimensions of the well-known result in three spatial dimensions. In
Section \ref{sec:toroidalsurface} we discuss equilibrium torus-like configurations of rotating fluids held
together by surface tension, and show that these share common features with their vacuum general-relativistic
counterpart, the black rings. In Section \ref{sec:stabilitysurface} we study the stability of the equilibrium
figures discussed in Section \ref{app:surfacetensiondrop} against small perturbations. We compute neutral modes
of oscillation and the onset of instability which allows one to derive the bifurcation diagram for the case of
one non-vanishing angular momentum. We finish this part with Section \ref{sec:cylindersurface} by reviewing the
Rayleigh-Plateau instability of a thin long cylinder in higher dimensions \cite{cardosodiasglrp}, which could
be a good model for the Gregory-Laflamme instability. We discuss possible effects of rotation and magnetic
fields.

In the second part we deal with self-gravitating fluids in higher dimensions. We start in Section
\ref{sec:cylindergravity} by considering an infinite cylinder of self-gravitating homogeneous fluid. We show
that this cylinder is unstable against a classical instability, here termed Dyson-Chandrasekhar-Fermi (DCF)
instability, which we demonstrate to be the exact analogue of the Gregory-Laflamme \cite{gl} instability of
black strings. We discuss implications of this result for rotating and charged fluids and black strings. We
then discuss in Section \ref{sec:torusselfgravitating} toroidal equilibrium configurations of self-gravitating
fluid (often termed Dyson rings) and we compare them to exact general relativistic toroidal-like
configurations, the black ring solution. Finally, in Section \ref{sec:MacLaurin} we write down the basic
equations that generalize the MacLaurin and Jacobi sequence to higher dimensions, equations which could be used
in future numerical work to reconstruct the whole bifurcation diagram. Oscillation modes of non-rotating globes
are derived in Appendix \ref{app:gravitysphere} and constitute a generalization to arbitrary number of
dimensions of the well-known result by Kelvin in three spatial dimensions. We close in Section
\ref{sec:conclusions} with a discussion of the results, implications and suggestions for future work.

This work is by no means complete. Many important topics are left for future work, we discuss some of these in
the concluding remarks.
\section*{Notation}
There are many different symbols used throughout the text. For clarity
we find it useful to make a short list of the most often used symbols.

\begin{itemize}
\item[] $D \rightarrow$ total number of spatial dimensions.

\item[] $x^i=(x^a,x^A) \rightarrow$ coordinates of a rotating body. We consider two particular cases:
in case I there is a single rotation plane, and $a=1,2$; $A=3,...,D$; in case II, $D$ is odd and there are
$\frac{D-1}{2}$ rotation planes. In this case $a=1,...D-1$; $A=D$.

\item[] $C_D \rightarrow$ the area of the unit $(D-1)$-sphere,
$C_D=\frac{2\pi^{D/2}}{\Gamma\left(\frac{D}{2}\right)}$.

\item[] $\varpi \rightarrow$ radial coordinate in cylindrical
coordinates.

\item[] $\omega\rightarrow$ Fourier transform variable. The time
dependence of any field is written as $e^{\omega t}$. In the case of a stable oscillation, it is related to the
frequency by $\omega^2 =-\sigma^2$. Note that pure imaginary values of $\omega$ mean a stable oscillation.

\item[] $\Omega\rightarrow$ angular velocity.

\item[] ${\cal B}({\bf x})\rightarrow$ Gravitational potential.

\item[] $G \rightarrow$ Newton's constant in $D-$dimensions, defined
by writing the gravitational potential as
\begin{equation}
{\cal B}({\bf x})\equiv G
\int_V\frac{\rho(\vec x')}{|\vec x-\vec x'|^{D-2}}d^Dx'\,.
\end{equation}
This is perhaps not the most common definition. A popular definition of Newton's constant is the one adopted by
Myers and Perry \cite{myersperry} and the relation between the two is derived in Appendix \ref{app:newton}.

\item[] $T\rightarrow$ magnitude of surface tension, defined by
$P=Tn^i_{,i} $ where $P$ is the pressure in the interior of the body
(we take the exterior pressure to be zero). We take ${\bf n}$ to be
the outward normal to the surface the body.

\item[] $P\rightarrow$ Pressure.

\item[] $V\rightarrow$ Volume.

\item[] $\rho\rightarrow$ density.

\item[] $\lambda \rightarrow$ wavelength of a given perturbation.

\item[] $k \rightarrow$ wavenumber $\equiv 2\pi/\lambda$.

\item[] $I^{ij}\equiv \int_V \rho x^ix^jd^Dx \rightarrow$ Moment
of inertia.
\end{itemize}
\part{Fluids held by surface tension}

\vskip 1cm

\section{\label{sec:symm}Symmetries and geometrical description of a multi-dimensional
rotating body}
Let us consider a stationary rotating object in $D$-dimensional
Euclidean space, with coordinates $x^i=(x^1,\dots,x^D)$. The rotation
group in $D$ dimensions, $SO(D)$, has $D(D-1)/2$ generators
$J^{ij}=J^{[ij]}$.  Each generator $J^{ij}$ corresponds to an $SO(2)$
rotation, on the plane $x^i-x^j$.

In a suitable coordinate frame $J=(J^{ij})$ can be brought to the standard form (see for instance
\cite{myersperry})
\begin{equation} J=\left(\begin{array}{ccc}
\begin{array}{cc} 0 & J_1 \\ -J_1 & 0 \\ \end{array} & & \\
& \begin{array}{cc} 0 & J_2 \\ -J_2 & 0 \\ \end{array} & \\
& & \ddots \\ \end{array} \right)\,. \end{equation}
This means that in this frame the rotation is a composition of a
rotation in the plane $x^1-x^2$, a rotation in the plane $x^3-x^4$,
and so on, without mixings. The rotation is then characterized by
$[D/2]$ generators (where $[\cdot]$ denotes the integer part), which
are the generators of the Cartan subalgebra of $SO(D)$ (namely, the
maximal abelian subalgebra of $SO(D)$).  In other words, in $D$
dimensions there are \begin{equation} {\cal N}=\left[\frac{D}{2}\right] \end{equation}
independent angular momenta.

We will consider two particular cases:
\begin{itemize}
\item[(I)] Only one angular momentum is non-vanishing: $J_1\neq 0$,
$J_{i>1}=0$.
\item[(II)] All angular momenta are non-vanishing, and they are equal:
$J_1=J_2=\dots=J_{\cal N}$; in this case we also assume $D$ to be odd.
\end{itemize}
These will be referred to as case I and II respectively throughout the
whole article.
\subsection{Case I. One non-vanishing angular momentum}
In this case we are selecting a one-dimensional rotation subgroup
\begin{equation} SO(2)\subset SO(D)\,. \end{equation}
This means selecting a single rotation plane, $x^1\!-\!x^2$.  We
assume our object has the maximal symmetry compatible with this
rotation:
\begin{equation} G=SO(2)\times SO(D-2)\subset SO(D)\label{sym}\,. \end{equation}
If we decompose the coordinates under $G$ as
\begin{equation} x^i=(x^a,x^A);~~~~a=1,2;~A=3,\dots,D\,,\label{coordinatesa} \end{equation}
and we define
\begin{eqnarray}
\varpi&\equiv&\sqrt{(x^1)^2+(x^2)^2}\nonumber\,,\\
z&\equiv&\sqrt{(x^3)^2+\dots+(x^D)^2}\,,\label{deftoz1} \end{eqnarray}
the surface of our object depends only on $\varpi$ and $z$, and its
equation can be written as
\begin{equation} z=f(\varpi) \,.\end{equation}
i.e.
\begin{equation} \Xi \equiv z-f(\varpi)=0\,, \end{equation}
therefore the surface does not depend on the angle in the $x^1\!-\!x^2$ plane, and it does not depend on the
$D-3$ angles among $x^3,\dots,x^D$. We also define
\begin{equation} \phi\equiv\frac{df}{d\varpi}\,. \end{equation}
Notice that it must be $\phi=-\infty$ when $z=0$ for the surface to be smooth.

Being $\varpi_{,a}=\frac{x^a}{\varpi}$, the normal to the surface is
\begin{equation} \Sigma_{,i}=(\Sigma_{,a},\Sigma_{,A})=\left(-\phi\frac{x^a}{\varpi},
\frac{x^A}{z}\right) \end{equation}
with modulus $\delta^{ij}\Sigma_{,i}\Sigma_{,j}=1+\phi^2$, so the
normal unit vector is
\begin{equation} n^i=\left(-\phi\frac{x^a}{\varpi\sqrt{1+\phi^2}},
\frac{x^A}{z\sqrt{1+\phi^2}}\right)\,. \end{equation}
Since the pressure is related to surface tension via $P=T n^i_{,i}$, we will need a general relation to compute
$n^i_{,i}$. Let us then compute $n^i_{,i}=n^a_{,a}+n^A_{,A}$ for this case I. We have
\begin{eqnarray} n^a_{,a}&=&-\frac{\phi'}{(1+\phi^2)^{3/2}}
-\frac{\phi}{\varpi\sqrt{1+\phi^2}}=-\frac{1}{\varpi}\frac{d}{d\varpi}
\frac{\varpi\phi}{\sqrt{1+\phi^2}}\nonumber\\
n^A_{,A}&=&\frac{1}{\sqrt{1+\phi^2}}\frac{D-3}{z}\,. \label{divna} \end{eqnarray}
%
\subsection{Case II. All angular momenta coincident, and $D$ odd.}
In this case the rotation group is broken to its Cartan subalgebra
\begin{equation} (SO(2))^{\frac{D-1}{2}}\subset SO(D)\,. \end{equation}
The rotation planes are
\begin{equation} x^1-x^2,~x^3-x^4,~\dots,~x^{D-2}-x^{D-1} \,,\end{equation}
and the $x^D$ axis is left unchanged by the rotation\footnote{Black
hole solutions with similar angular momentum configurations have been
considered in \cite{kunz}.}. The decomposition of the coordinates is
now
\begin{equation} x^i=(x^a,x^A);~~~a=1,\dots,D-1\,;~~~A=D \,.\label{coordinatesb} \end{equation}
We define
\begin{eqnarray}
\varpi&\equiv&\sqrt{(x^1)^2+(x^2)^2+\dots+(x^{D-1})^2}\nonumber\,,\\
z&\equiv&x^D\,. \end{eqnarray}
The surface does not depend on the $D-2$ angles among the $x^a$. We also define
\begin{equation} \phi\equiv\frac{df}{d\varpi}\,. \end{equation}
Notice that it must be $\phi=-\infty$ when $z=0$ for the surface to be smooth.

Being $\varpi_{,a}=\frac{x^a}{\varpi}$, the normal to the surface is
\begin{equation} \Sigma_{,i}=\left(-\phi\frac{x^a}{\varpi},0\right)\,, \end{equation}
with modulus $\delta^{ij}\Sigma_{,i}\Sigma_{,j}=1+\phi^2$, so the
normal unit vector is
\begin{equation} n^i=\left(-\phi\frac{x^a}{\varpi\sqrt{1+\phi^2}},0\right)\,. \end{equation}
Finally,
\begin{eqnarray} n^i_{,i}&=&n^a_{,a}=-\frac{1}{\varpi^{D-2}}\frac{d}{d\varpi}\left
(\frac{\varpi^{D-2}\phi}{\sqrt{1+\phi^2}}\right)\,.\label{divnb} \end{eqnarray}
\section{\label{sec:hydroeqs}Hydrodynamical and hydrostatical equations}
Let ${\bf X}=(X^i)$ the position vector of a fluid element in the inertial frame, and ${\bf x}=(x^i)$ its
position vector in the rotating frame, i.e.
\begin{equation} {\bf X}=R{\bf x} \end{equation}
where $R=R_I$ or $R_{II}$ according to the case under consideration and
\begin{equation} R_I=\left(\begin{array}{cc|ccc}
\cos\Omega t & \sin\Omega t &&& \\
-\sin\Omega t & \cos\Omega t &&& \\
\hline
& & 1 && \\
& & & \dots & \\
& & & & 1 \\
\end{array}\right)\,;
~~~~~~~~ R_{II}=\left(\begin{array}{cccc}
\begin{array}{cc} \cos\Omega t & \sin\Omega t \\
-\sin\Omega t & \cos\Omega t \\ \end{array} &   &   &   \\ 
&   \ddots    &   \\ 
&  &  \begin{array}{cc} \cos\Omega t & \sin\Omega t \\
-\sin\Omega t & \cos\Omega t \\ \end{array}   &   \\ 
&   &   &   1 \\
\end{array}\right)
\end{equation}
are $D$-dimensional rotation matrices in the cases I, II respectively. We introduce the velocity ${\bf \cal V}$
and acceleration ${\bf \cal A}$ by
\begin{eqnarray}
\dot{\bf X}&=&R{\bf  \cal V}\nonumber\\
\ddot{\bf X}&=&R{\bf \cal A} \end{eqnarray} where an overdot denotes time differentiation.

The hydrodynamical equation, in the inertial frame, is
\begin{equation} \ddot X^i=-\frac{1}{\rho}\frac{\partial P}{\partial X^i}\label{eqmotion} \end{equation}
(with $P$ pressure and $\rho$ constant density of the fluid) which,
applying $R^{-1}$, gives
\begin{equation} {\cal A}^i=-\frac{1}{\rho}\frac{\partial}{\partial x^i}P\,. \end{equation}
Let us compute ${\bf \cal A}$. We have
\begin{equation} {\bf \cal V}=R^T\dot R{\bf x}+\dot{\bf x} \end{equation}
and then
\begin{equation} {\bf \cal A}=R^T\dot R{\bf \cal V}+\dot{\bf \cal V}= (R^T\dot R)^2{\bf x}+2R^T\dot R \dot{\bf x}+\ddot{\bf
x}\,. \end{equation}
We have that
the only non-vanishing components of $R^T\dot R$ are
\begin{equation} (R^T\dot R)^{ab}=-\Omega\epsilon^{ab} \end{equation}
where $\epsilon^{ab}=\epsilon^{[ab]}$, and $\epsilon^{12}=1$ in the case I,
$\epsilon^{12}=\epsilon^{34}=\dots=\epsilon^{(D-2)(D-1)}=1$ in the case II.

Therefore, in both cases $((R^T\dot R)^2)^{ab}=\Omega^2\delta^{ab}$, and
\begin{eqnarray} \ddot x^a&=&-\frac{1}{\rho}\frac{\partial P}{\partial x^a}+\Omega^2
x^a+2\Omega\epsilon^{ab}\dot x^b \label{eq11}\\
\ddot x^A&=&-\frac{1}{\rho}\frac{\partial P}{\partial x^A}\,.\label{hydyn} \end{eqnarray}
The second term in the right hand side of (\ref{eq11}) is the centrifugal force and the third term is the
generalized Coriolis force. At equilibrium $\dot x^i=\ddot x^i=0$, therefore
\begin{eqnarray}
\frac{\partial P}{\partial x^a}&=&\rho\Omega^2 x^a\nonumber\,,\\
\frac{\partial P}{\partial x^A}&=&0 \,.\end{eqnarray}
thus 
\begin{equation} P=P_0+\frac{1}{2}\rho\Omega^2\varpi^2\,. \label{press} \end{equation}
This formula is valid both in case I and in case II, i.e., when there is one non-vanishing angular momentum and
when all of them coincide (and $D$ is odd).

\section{\label{sec:perturbations}The virial theorem, surface-energy tensor and small perturbations}
We would ultimately wish to assess stability of rotating drops, in order to generalize Fig. \ref{fig:brown} to
higher dimensions. This calculation would also provide some hints as to whether or not one should expect to see
instabilities in black objects in higher dimensions, as we stressed in the Introduction. To attack this problem
we will follow Chandrasekhar's \cite{chandra} approach, to which we refer for further details. Alternative
(numerical) approaches, perhaps more complete but more complex also, are described by Brown and Scriven
\cite{brownscriven} and also Heine \cite{heine}.
\subsection{The virial theorem in the inertial frame}
Consider a fluid of volume $V$ with uniform density $\rho$, bounded by
a surface $S$. The external pressure, for simplicity, is taken to be
zero. Due to the existence of surface tension, the pressure $P$
immediately adjacent to $S$, on the interior, is given in terms of the
surface tension magnitude $T$ by
\begin{equation} P=Tn^i_{,i} \label{eqdrop}\,. \end{equation}
Multiplying (\ref{eqmotion}) by $X^j$ and integrating over the volume $V$ of the fluid we get
\begin{equation} \frac{d}{dt}\int_V \rho \dot{X}^iX^j d^DX=2{\cal I}^{ij}-\int_V X^j \frac{\partial P}{\partial X^i} d^DX
\,,\label{st1} \end{equation}
where ${\cal I}^{ij}=\frac{1}{2}\int_V \rho \dot{X}^i \dot{X}^j d^D X$ is the inertia tensor. Integrating by
parts the second term on the R.H.S. of (\ref{st1}) we have
\begin{equation} \int_V X^j \frac{\partial P}{\partial X^i} d^DX=\int_S X^jPdS^i -\delta^{ij}\int_V P d^DX\,,\label{st2} \end{equation}
where $dS^i=n^idS$ and $dS$ is the $(D-1)$-dimensional surface element
on $S$. Using (\ref{eqdrop}) we further have
\begin{equation} \int_V X^j \frac{\partial P}{\partial X^i} d^DX= T\int_S X^jn^i_{,i} dS^i-\delta^{ij} \int _V P
d^DX\,.\label{st3} \end{equation}
Introducing this result in (\ref{st1}) one gets the virial theorem
\begin{equation} \frac{d}{dt}\int_V \rho \dot{X}^iX^j d^DX=2{\cal I}^{ij}+ {\mathfrak S}^{ij}+\delta^{ij}\Pi \,,\end{equation}
where
\begin{eqnarray} \Pi&=&\int_V P d^DX\,,\\
 {\mathfrak S}^{ij}&=&-T\int_S X^j n^k_{,k}dS^i\,. \end{eqnarray}
The quantity ${\mathfrak S}^{ij}$ is the surface-energy tensor \cite{chandra}. Conservation of angular momentum
guarantees that $\mathfrak{S}^{ij}$ is symmetric. Indeed,
\begin{equation}
\frac{d}{dt}\int_V\rho(\dot X^iX^j-\dot X^jX^i)d^D X=\mathfrak{S}^{ij}-\mathfrak{S}^{ji}=0\,.
\end{equation}
\subsection{The virial theorem in the rotating frame}
In a rotating frame of reference the equations of motion are written down in (\ref{hydyn}). Multiplying those
equations by $x^j$, integrating over $V$ and using the previous sequence of transformations we get
\begin{eqnarray} \frac{d}{dt}\int_V \rho \dot{x}^a x^j d^Dx&=&2{\cal I}^{aj}+{\mathfrak
S}^{aj}+\delta^{aj}\Pi+\Omega^2I^{aj}
+2\Omega \epsilon^{ac}\int_V \rho \dot{x}^{c}x^{j}d^Dx \,,\label{v1}\\
 \frac{d}{dt}\int_V \rho \dot{x}^A x^j d^Dx&=&2{\cal I}^{Aj}+
{\mathfrak S}^{Aj}+\delta^{Aj}\Pi \,,\label{v2}\end{eqnarray}
where
\begin{equation} I^{aj}=\int_V \rho x^{a}x^{j}d^Dx\,. \end{equation}
The tensor $\mathfrak{S}^{ij}$ is symmetric in the rotating frame,
too, as can be shown by transforming it with the rotation matrix $R$.

We remind the reader that according to our conventions the indices $i,j$ can take any values between $1$ and
$D$, but the indices $a,\,b$ and $A,\,B$ are restricted according to the situation we consider (case I or II),
described in Equations (\ref{coordinatesa}) and (\ref{coordinatesb}). In the following we always consider the
rotating reference frame.
\subsection{\label{virial} Virial equations for small perturbations}
Having solved for the equilibrium equations and found the shape for a given rotating drop (this problem will be
explicitly worked out in the next Section \ref{sec:eqshapesurface}), we ask whether this is a stable
equilibrium configuration or not. To answer this we follow the usual procedure to assess stability: we perturb
the equilibrium configuration. Suppose then that a drop in a state of equilibrium is slightly disturbed. All
background quantities $Q_0$ are then changed by an amount $\delta Q_0$. The motions are described by a
Lagrangian displacement of the form
\begin{equation} {\bf \xi}(x)e^{\omega t}\,, \end{equation}
where $\omega$ is to be determined from the problem. In the case of constant density, $\xi$ is divergence free.
Define
\begin{equation} \Gamma^{i;j}=\int_V \rho \xi^ix^jd^Dx\,, \end{equation}
where the semi-colon indicates that a moment with respect to
coordinate space is being evaluated. Define also
\begin{equation} \Gamma^{ij}\equiv  \Gamma^{i;j}+ \Gamma^{j;i}\,. \end{equation}
Then the perturbation equations read \cite{chandra}
\begin{eqnarray} \omega^2\Gamma^{a;j}-2\omega\Omega \epsilon^{ac}\Gamma^{c;j} &=&\delta {\mathfrak S}^{aj}+
\Omega^2\Gamma^{aj}+\delta^{aj}\delta \Pi \,,\label{eqvir1}\\
\omega^2\Gamma^{A;j}&=&\delta {\mathfrak S}^{Aj}+\delta^{Aj} \delta \Pi \,.\label{eqvir2} \end{eqnarray}
The surface integrals $\delta\mathfrak S^{ij}$ have the form
\begin{eqnarray}
\delta\mathfrak S^{ij}&=&-T\delta\int_Sx^jn^k_{~,k}dS^i
=-T\left[\int_S\xi^jn^k_{~,k}dS^i+\int_Sx^j\delta(n^k_{~,k})dS^i
+\int_Sx^jn^k_{~,k}\delta(dS^i)\right]\nonumber\\
&=&-T\left[\int_S\xi^jn^k_{~,k}dS^i+\int_Sx^j\delta(n^k_{~,k})dS^i
-\int_Sx^jn^k_{~,k}\frac{\partial \xi^l}{\partial x^i}dS^l\right]\label{eqkS}\\
\end{eqnarray}
where we have used the general result (see for instance \cite{chandra})
\begin{equation}
\delta(dS^i)=\xi^j_{~,j}dS^i-\frac{\partial \xi^j}{\partial x^i}dS^j
\end{equation}
and $\xi^j_{~,j}=0$. Furthermore, we have (see \cite{chandra})
\begin{eqnarray}
\delta(n^i_{~,i})=(\delta n^i)_{~,i}-\frac{\partial n^i}{\partial x^k}
\frac{\partial \xi^k}{\partial x^i}\nonumber\\
\delta n^i=\left(n^kn^l\frac{\partial \xi^k}{\partial x^l}\right)n^i -n^j
\frac{\partial \xi^j}{\partial x^i}\,.\label{expdiv}
\end{eqnarray}

Equations (\ref{eqvir1}), (\ref{eqvir2}) can be decomposed as
\begin{eqnarray} \omega^2\Gamma^{a;b}-2\omega\Omega \epsilon^{ac}\Gamma^{c;b}&=&\delta {\mathfrak S}^{ab}+
\Omega^2\Gamma^{ab}+\delta^{ab}\delta \Pi \,,\label{even1}\\
\omega^2\Gamma^{A;B}&=&\delta {\mathfrak S}^{AB}+
\delta^{AB}\delta\Pi \,,\label{even2}\\
\omega^2\Gamma^{a;A}-2\omega\Omega \epsilon^{ac}\Gamma^{c;A}&=&\delta {\mathfrak S}^{aA}+
\Omega^2\Gamma^{aA} \,,\label{odd1}\\
\omega^2\Gamma^{A;a}&=&\delta {\mathfrak S}^{Aa} \label{odd2}\,.\end{eqnarray}
These are the basic equations we will use later on to perform a stability analysis of rotating drops in higher
dimensions. To complete the calculations we first need to compute the equilibrium figures of rotating drops and
we this in the following.
\section{\label{sec:eqshapesurface}Equilibrium shape of a rotating drop}

\subsection{\label{sec:eqcasea} Case I. One non-vanishing angular momentum}
From equations (\ref{divna}), (\ref{press}), (\ref{eqdrop}) we find
\begin{equation}
P_0+\frac{1}{2}\rho\Omega^2\varpi^2=T\left(-\frac{1}{\varpi}\frac{d}{d\varpi}
\frac{\varpi\phi}{\sqrt{1+\phi^2}}
+\frac{1}{\sqrt{1+\phi^2}}\frac{D-3}{z}\right)\,,\label{equationcasea}
\end{equation}
thus
\begin{equation} \frac{1}{2}P_0\varpi^2+\frac{1}{8}\rho\Omega^2\varpi^4=
 -T\frac{\varpi\phi}{\sqrt{1+\phi^2}}+(D-3)TI+C\label{eq0}
\end{equation}
where $C$ is a constant, and $I$ is the integral
\begin{equation} I(\varpi)=\int_0^\varpi{\frac{d\varpi'\varpi'}{f(\varpi')\sqrt{1+ \left(\frac{df}{d\varpi'}
\right)^2}}}\,.\label{I} \end{equation}
If the drop encloses the origin (which we suppose for the time being),
then $\varpi=0$ must belong to the surface; being $I(0)=0$, this means
that $C=0$ and equation (\ref{eq0}) reduces to
\begin{equation} \frac{1}{2}P_0\varpi^2+\frac{1}{8}\rho\Omega^2\varpi^4=
-T\frac{\varpi\phi}{\sqrt{1+\phi^2}}+(D-3)TI\,.\label{eq}
\end{equation}
When $D=3$ this equation can be solved analytically \cite{chandra}, but when $D>3$ this is not possible,
because equation (\ref{eq}) cannot be expressed in the form $\frac{df}{d\varpi}=F(\varpi)$.

We thus have to numerically integrate equation (\ref{equationcasea}), which we re-write as
\begin{equation}
\frac{P_0}{T}+\frac{1}{2T}\rho\Omega^2\varpi^2=\left(-\frac{1}{\varpi}\frac{d}{d\varpi}
\frac{\varpi\phi}{\sqrt{1+\phi^2}}
+\frac{1}{\sqrt{1+\phi^2}}\frac{D-3}{z}\right)\,,\label{equationcasea2}
\end{equation}
We can re-scale coordinates $\varpi \rightarrow a\varpi$ and $z
\rightarrow az$, where $a$ will henceforth be the equatorial radius of
the equilibrium figure. In this way all quantities are dimensionless
and equation (\ref{equationcasea2}) is written as
\begin{equation} \frac{P_0a}{T}+4\Sigma_a\varpi^2=-\frac{1}{\varpi}\frac{d}{d\varpi}
\frac{\varpi\phi}{\sqrt{1+\phi^2}}
+\frac{1}{\sqrt{1+\phi^2}}\frac{D-3}{z}\,,\label{equationcasea3} \end{equation}
with $\Sigma_a \equiv\frac{\rho \Omega^2a^3}{8T}$.

This equation can be integrated numerically, starting at $\varpi=0$
where $z=z(0)$ and $z'(0)\equiv \phi=0$. In this sense we are solving
an eigenvalue equation: given $\Sigma_a$, $P_0a/T$ and $z(0)$ are
determined by requiring that at the equator $r=a$ we have $z(a)=0$ and
$z'(a)=-\infty$. The problem is highly non-linear, so the numerical
procedure involved varying {\it both} $P_0a/T$ and $z(0)$
simultaneously over a rectangular grid, in order to determine those
values for which the outer boundary conditions ($z(a)=0$ and
$z'(a)=-\infty$) were satisfied. The accuracy gets worse for higher
$\Sigma$ and $D$ but one usually has a precision of about $5\%$ in our
numerical results.
\begin{figure}[ht]
\centerline{\includegraphics[width=8 cm,height=8 cm] {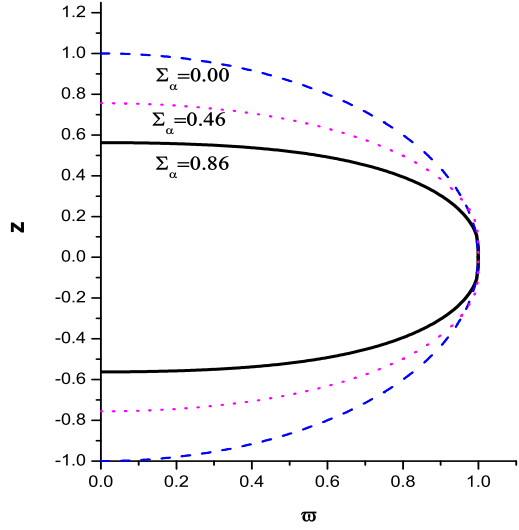}} \caption{Typical (axi-symmetric) equilibrium
figures of rotating drops for case I (one non-vanishing angular momentum), in $D=4$.  The (blue) dashed line
corresponds to a spherical surface with $\Sigma_a=0$. The (magenta) dotted line corresponds to $\Sigma_a=0.46$
and the (black) solid line corresponds to the limit $\Sigma_a=0.86$. In all three cases the equatorial radius
is unity.} \label{fig:eqfigsa}
\end{figure}
Numerical results for $D=4$ are shown in Figure \ref{fig:eqfigsa} for
some values of $\Sigma_s$, which can be compared to the $D=3$ results
presented by Chandrasekhar in \cite{chandra}. For $\Sigma_a=0$, the
non-rotating case, we start by verifying numerically that the sphere
is a solution, as it should, with $P_0a/T=D-1$. The equilibrium
figures are very similar and indeed the similarity seems to continue
for larger $D$.
The three different equilibrium shapes shown in Figure \ref{fig:eqfigsa} all enclose the origin. As in the
$D=3$ case (see \cite{chandra}), we find that for $\Sigma_a$ larger than a certain critical value, the drop no
longer enclosed the origin, giving rise to toroidal shapes. The condition to be satisfied at the critical
$\Sigma$ is that $z=0$ at $r=0$. For $D=3$ the critical value of $\Sigma_a$ is $\Sigma_a\sim 2.32911$. For
$D>3$ it is harder to determine the critical parameter, since there is no integral expression for the shape.
Nevertheless we determine $\Sigma_{\rm crit}\sim 2.3$ for $D=4,5$ with a $5\%$ uncertainty. An open important
problem is to determine if a critical point exists for all $D$. If not, this might mean that the bifurcation
curve \ref{fig:brown} does not bend down, and therefore that the angular momentum is unbounded, which might be
a good analogue for what happens with ultra-rotating \cite{myersperry} black holes. This may require more
accurate numerical integrations, since we cannot trust our method for $D>5$.
\subsection{Case II. All angular momenta coincident, and $D$ odd.}
In case II we will find that the results are very similar to those for $D=3$. From equations (\ref{divnb}),
(\ref{press}), (\ref{eqdrop}) we find
\begin{equation} P_0+\frac{1}{2}\rho\Omega^2\varpi^2=-\frac{T}{\varpi^{D-2}}
\frac{d} {d\varpi}\left
(\frac{\varpi^{D-2}\phi}{\sqrt{1+\phi^2}}\right )\,, \end{equation}
which is immediately integrated to give
\begin{equation} P_0\frac{\varpi^{D-1}}{D-1}+\frac{\rho
  \Omega^2\varpi^{D+1}}{2(D+1)}
=-\frac{T\varpi^{D-2} \phi}{\sqrt{1+\phi^2}}+C\,, \label{eq1} \end{equation}
where $C$ is a constant.

\begin{figure}[ht]
\centerline{\includegraphics[width=8 cm,height=10 cm] {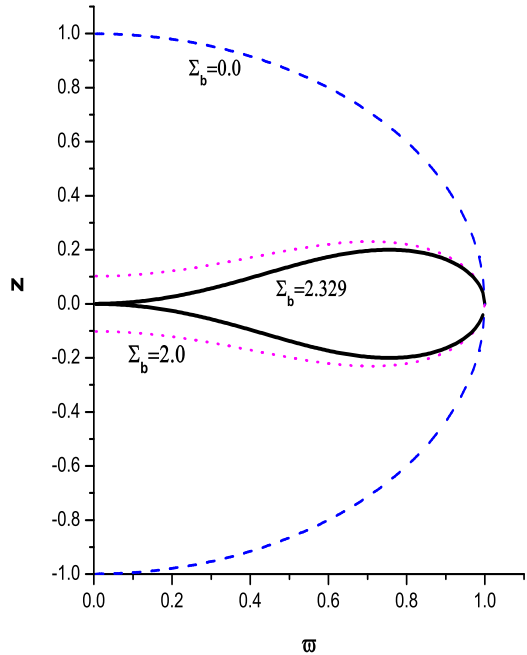}} \caption{Typical equilibrum figures of
rotating drops.  The (blue) dashed line corresponds to a spherical surface with $\Sigma_b=0$. The (magenta)
dotted line corresponds to $\Sigma_b=2$ and the (black) solid line corresponds to the limit $\Sigma_b=2.329$.
In all three cases the equatorial radius was chosen to be unity.} \label{fig:eqfigs}
\end{figure}

Let's first suppose the drop encloses the origin. In this case, there
are points where $\varpi=0$ and therefore we must have $C=0$. Equation
(\ref{eq1}) can then be written as
\begin{equation} P_0\frac{\varpi}{D-1}+\frac{\rho \Omega^2\varpi^3}{2(D+1)}
=-\frac{T\phi}{\sqrt{1+\phi^2}}\,. \label{eq2}
\end{equation}
At the equator, where $\varpi=a$ we must have $\phi \rightarrow
-\infty$ and then (\ref{eq1}) implies that
\begin{equation} \frac{P_0a}{(D-1)T}=1-\Sigma_b\,,\quad \Sigma_b \equiv
\frac{\rho \Omega^2a^3}{2(D+1)T}\,.\label{sigmadef}
\end{equation}
Equations (\ref{eq2})-(\ref{sigmadef}) are the same as equations (A4)-(A5) by Chandrasekhar \cite{chandra}. So,
figures of equilibrium enclosing the origin do not depend on the dimensionality of spacetime, if measured in
terms of the dimensionless number $\Sigma_b$ (by this we mean of course that $z(\varpi,\Sigma_b)$ is
independent of $D$; anyway, the relation between $\Sigma_b$ and $\Omega$ depends on $D$, thus
$z(\varpi,\Omega)$ depends on $D$). By measuring $\varpi$ and $\phi$ in units of $a$, we get
\begin{equation} \phi=\frac{d f}{d\varpi}=-\frac{\varpi\left (1- \Sigma_b+\Sigma_b
\varpi^2\right )}{\sqrt{1-\varpi^2\left
(1-\Sigma_b+\Sigma_b \varpi^2\right )^2}}\,, \end{equation}
which can be written as
\begin{equation} z=f(\varpi)=\int_{\varpi}^{1}\frac{w\left (1-\Sigma_b+\Sigma_b w^2
\right )}{\sqrt{1-w^2\left
(1-\Sigma_b+\Sigma_b w^2\right )^2}}d w\,. \end{equation}
Transforming to a new variable $\theta$ defined through
\begin{equation} \varpi^2=1-A\tan^2({\theta/2})\,,\quad A\equiv
\frac{\sqrt{1+2\Sigma_b}}{\Sigma_b}\,, \end{equation}
the integral representation can be simplified to
\begin{equation} z=\frac{1}{2\Sigma_b\sqrt{A}}\int_0^{\phi}\left(1+A\Sigma_b-
\frac{2A\Sigma_b}{1+\cos{\theta}}\right
)\frac{d\theta}{\sqrt{1-\chi^2\sin^2\theta}}\,, \label{zsol}\end{equation}
where
\begin{equation} \chi^2=\frac{1}{2}\left [1+\frac{1}{A}\left ( \frac{1}{2}+
\frac{1}{\Sigma_b} \right )\right ]\,. \end{equation}
Here, $\phi$ is related to the value of $\varpi$ at the boundary by
\begin{equation} \varpi=\sqrt{1-A\tan^2{\phi/2}} \end{equation}
The solution (\ref{zsol}) can be expressed in terms of elliptic integrals
\begin{equation}
F(\chi,\phi)=\int_0^{\phi}\frac{d\theta}{\sqrt{1-\chi^2\sin^2\theta}}
\,,\quad E(\chi,\phi)=\int_0^{\varpi}
d\theta \sqrt{1-\chi^2\sin^2\theta}\,, \end{equation}
as
\begin{equation} z=\frac{1}{2\Sigma_b \sqrt{A}}\left [\left (1-\Sigma_b A\right
  )F(\chi,\phi) + 2A\Sigma_b \left (E(\chi,\phi) -\frac{\sin\phi
  \sqrt{1-\chi^2\sin^2\phi}}{1+\cos \phi}\right )\right ]
\label{shape}\,.\end{equation}
Some representative equilibrium solutions are displayed in Figure \ref{fig:eqfigs}, for some values of
$\Sigma_b$. The shapes are very similar to their ``case I'' relatives. We show typical shapes for high values
of $\Sigma$ (these can be compared with the ones presented in Figure 1 of Chandrasekhar \cite{chandra}).

We have been dealing with drops that enclose the origin, and found
that a closed expression for the shape is given by (\ref{shape}). When
does this sequence end? A necessary and sufficient condition for the
drop to enclose the origin is that $z=0$ and $\xi=0$ are
simultaneously satisfied. Now, the condition $\xi=0$ implies that
$\phi=\phi_{\rm max}=2\arctan A^{-1/2}$. Substituting this value in
(\ref{shape}) we get
\begin{equation} \Sigma_{\rm max}=2.32911\,. \end{equation}
For values of $\Sigma_b$ larger than the above, the figure no longer
encloses the origin.
%
\section{\label{sec:toroidalsurface}Equilibrium toroidal shapes}
In the general case of figures which do not enclose the origin (i.e.,$\varpi=0$ is not attained), we have to
use equation (\ref{equationcasea}) and (\ref{eq1}) for cases I and II respectively. Here we shall focus on case
II, equation (\ref{eq1}), since it is easily manageable. In three spatial dimensions, solutions to equation
(\ref{eq1}) include ellipsoids, disks, biconcave and torus-like (or ``wheels'' in the terminology of
\cite{rotatingdropexperiments2}) forms \cite{heine,rotatingdropexperiments2}. We now wish to consider if one
can use these toroidal fluid configuration to model strong gravity configurations with the same topology. In
particular, we consider a torus with a large radius $R$ and small thickness $r_+$ (the reason being that it is
in this regime that one really expects a ``pure'' torus).

\begin{figure}[ht]
\centerline{\includegraphics[width=8.5 cm,height=6 cm] {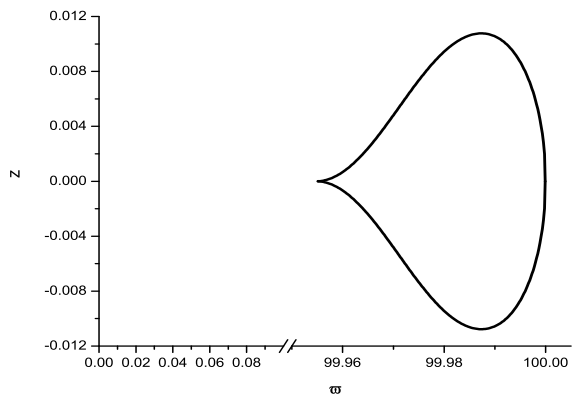}} \caption{Typical ``smooth'' Type A
torus-like equilibrium shape.  We have taken $D=4$ and $\Lambda=3.5,\,R_1=99.9549,\,R_2=100$.}
\label{fig:smoothtorus}
\end{figure}

We will first look for torus-like solutions which start smoothly (i.e. $\phi=0$) at some distance $R_1$. We
will call these ``type A'' toroidal solutions. The outer radius will be denoted $R_2$, where $\phi=-\infty$. In
this case, we can easily determine the constants $p_0\,,C$ and write (\ref{eq1}) as
\begin{equation} P_0\frac{\varpi^{D-1}-R_1^{d-1}}{T(D-1)}+\frac{\rho \Omega^2 \left
(\varpi^{D+1}-R_1^{D+1}\right
)}{2T(D+1)}=-\frac{\varpi^{D-2}\phi}{\sqrt{1+\phi^2}}\,, \label{eq122}
\end{equation}
where
\begin{equation} P_0\frac{R_2^{D-1}-R_1^{D-1}}{T(D-1)}=R_2^{D-2}-\frac{\rho \Omega^2 \left
(R_2^{D+1}-R_1^{D+1}\right)}{2T(D+1)}\,. \label{eq1222} \end{equation}
We have numerically integrated equation (\ref{eq1222}) for several values of the parameters, and we find the
law
\begin{equation} R_2 \left (R_2-R_1 \right )^2(D+1)\Lambda\simeq 3.57\,, \end{equation}
which is independent of dimensionality $D$. Here
\begin{equation} \Lambda\equiv \frac{\rho \Omega^2}{2T(D+1)}\,.\label{lambdadef} \end{equation}
The meridional cross-section for a typical type A torus is displayed in Fig. \ref{fig:smoothtorus}.
As shown by Gulliver \cite{gulliver} and Smith and Ross
\cite{smithross} there are several possible families of toroidal
fluids (some of which have been experimentally observed already, see
\cite{rotatingdropexperiments2}).  We have just described one of those
families, that starts smoothly at $R_1$.

\begin{figure}[ht]
\centerline{\includegraphics[width=8.5 cm,height=6 cm] {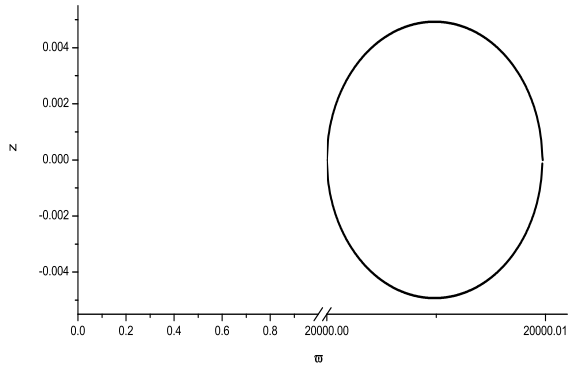}} \caption{Typical Type B torus-like
equilibrium shape. We have taken $D=4$ and $\Lambda=10^{-7},\,R_1=2\times 10^4,\,R_2=2.00000099\times 10^4$. We
chose $\frac{P_0}{T(D-1)}=1$.} \label{fig:torus}
\end{figure}

We can find a second family, the type B toroidal shape, which is more similar to the gravitational Dyson rings
which will be described in Section \ref{sec:torusselfgravitating}. What we now look for is a family which
starts ``abruptly'' at $R_1$. Our numerical experiments seem to suggest a very rich behavior for the members of
this family. For very small $\Lambda$, we find the following approximate scaling law,
\begin{equation} r_+ R^2\Lambda \sim \frac{1.96}{D+1}\,,\quad R\equiv R_2\sim
R_1\,.\label{typeIItorus} \end{equation}
and the meridional cross-section of a typical member of this family is
shown in Figure \ref{fig:torus}. Such a family seems to have been
discussed already by Appell \cite{appell}.
Let's pause a moment on this scaling law. If, following the suggestion in \cite{cardosodiasglrp}, we try to
model gravitating black bodies with fluids held together by surface tension, the first law suggests
\cite{cardosodiasglrp} that the surface tension is the equivalent of the Hawking temperature (up to a
multiplicative constant). Now, the temperature of a black hole scales as $1/r$, where $r$ is some typical
radius. For large ``black rings'' the relevant radius is the thickness $r_+$ of the ring. Thus we set $T \sim
1/r_+$. Using (\ref{lambdadef}) and (\ref{typeIItorus}) we get
\begin{equation} R^2\Omega^2\frac{M}{R r_+^{D-3}}\sim 1\,, \end{equation}
where we replaced $\rho=M/V$.

On the other hand, the mass per unit length of such a ``black torus'' should be given by the mass of a $D-1$
Tangherlini black hole \cite{tangherlini} (i.e., a higher dimensional non-rotating black hole):
\begin{equation} (D-2)C_DG M \sim 4\pi(D-2)C_{D-2}R\,r_+^{D-3}\,. \label{massblackobjects0} \end{equation}
Using the ``constitutive'' relation (\ref{massblackobjects0}) for black objects  and $J\sim \Omega M R$ we get
finally
\begin{equation} \frac{J^2}{M^2R^2} \sim 1\,. \label{surfacering}\end{equation}
Fortunately, there is an exact solution of Einstein's equations
describing a general relativistic black ring, in five dimensions
($D=4$) \cite{Emparan:2001wn}: the black ring solution by Emparan and
Reall \cite{Emparan:2001wn}. The black ring solution is described by
two parameters $\nu$ and $R$. They are characterized by having mass
and angular momentum equal to \cite{Emparan:2001wn,Elvang:2003mj}:
\begin{eqnarray} M&=&\frac{3\pi
R^2}{4G_L}\frac{\bar\lambda(1+\bar\lambda)}{1+\nu}\,\\ J&=&\frac{\pi
R^3}{2G_L}\frac{\sqrt{\bar\lambda \nu}(1
+\bar\lambda)^{5/2}}{(1+\nu)^2}\,. \end{eqnarray}
Here, the parameter $\bar\lambda=2\nu/(1+\nu^2)$ for black
rings. Large rings can be obtained by setting $\nu \rightarrow 0$, in
which case $R$ can be looked at as the radius of the ring. For very
small $\nu$ we then have
\begin{equation} \frac{J}{MR}=\frac{\sqrt{2}}{3} \label{grring}\,, \end{equation}
This is consistent with relation (\ref{surfacering}) obtained in the model of a fluid held together by surface
tension. The case $D=3$ seems to be special for in three dimensions there are no black strings (all black holes
must be topological spherical in this case), whereas the surface tension model seems to allow for rings;
\section{\label{sec:stabilitysurface}Stability analysis of a rotating drop in case I. one
non-vanishing angular momentum}
To complete our discussion on drops held by surface tension, we consider the generalization to $D$-spatial
dimensions of Figure \ref{fig:brown}, the bifurcation and stability diagram. Here we restrict ourselves to case
I, i.e. only one non-vanishing angular momentum, corresponding to rotation in the plane $x^1-x^2$, and we will
necessarily leave many topics for future research. For instance, all asymmetric figures will be left out of
this exercise. We will follow the approach of Chandrasekhar \cite{chandra}, based on the formalism of virial
tensors. The special case in which the rotation is zero, i.e., a spherical drop, is considered in Appendix
\ref{app:surfacetensiondrop} where we compute the oscillation modes of a $D$-dimensional incompressible fluid
sphere held by surface tension.

We have found the basic equations describing the equilibrium shape of
rotating drops in the previous Sections.  We would now like to
investigate if these shapes are stable. The formalism to do this was
laid down in Section \ref{virial}, and equations
(\ref{even1})-(\ref{odd2}) are the most general set of equations
describing perturbed rotating drops.

From now on we will restrict our analysis to a special kind of perturbations, termed ``toroidal'' modes by
Chandrasekhar \cite{chandra}. The reason is that these modes are expected to play a role in the bifurcation and
stability analysis. Future work should include an analysis of the other modes (termed ``transverse-shear'' and
``pulsation modes''). We refer the reader to the work by Chandrasekhar \cite{chandra} for further details on
the formalism. The toroidal modes are defined by
\begin{eqnarray}
\xi^1&=&\alpha x^1+\beta x^2\nonumber\\
\xi^2&=&\beta x^1-\alpha x^2\nonumber\\
\xi^A&=&0\label{tor}
\end{eqnarray}
with $\alpha,\beta$ constant parameters.

As shown in the Appendix \ref{toroidal}, if a perturbation has the
form (\ref{tor}), the only non-vanishing virial integrals are
\begin{eqnarray}
\Gamma^{1;1}&=&-\Gamma^{2;2}=\pi\rho\alpha\frac{C_{D-2}}{D-2}
\int_0^a\varpi^3\left[f(\varpi)\right]^{D-2}d\varpi
\label{vir11}\\
\Gamma^{1;2}&=&\Gamma^{2;1}=\pi\rho\beta\frac{C_{D-2}}{D-2}
\int_0^a\varpi^3\left[f(\varpi)\right]^{D-2}d\varpi
\label{vir12}
\end{eqnarray}
where $\varpi=a$ is the intersection of the surface of the drop with the $x^1-x^2$ plane.  These quantities
satisfy the equations (\ref{even1}), which have the form
\begin{eqnarray} \omega^2\Gamma^{1;1}-2\omega\Omega \Gamma^{2;1} &=&\delta {\mathfrak{S}}^{11}+
\Omega^2\Gamma^{11}+\delta \Pi \,,\label{1e}\\
\omega^2\Gamma^{2;2}+2\omega\Omega\Gamma^{1;2}&=& \delta {\mathfrak{S}}^{22}+
\Omega^2\Gamma^{22}+\delta \Pi \,,\label{2e}\\
\omega^2\Gamma^{1;2}-2\omega\Omega \Gamma^{2;1}&=& \delta {\mathfrak{S}}^{12}+
\Omega^2\Gamma^{12} \,,\label{3e}\\
\omega^2\Gamma^{2;1}+2\omega\Omega\Gamma^{1;2}&=& \delta {\mathfrak{S}}^{21}+ \Omega^2\Gamma^{21}\label{4e}
\end{eqnarray} and, combined, become
\begin{eqnarray} \frac{1}{2}\omega^2\left(\Gamma^{11}-\Gamma^{22}\right) -2\omega\Omega \Gamma^{12}&=&\delta
{\mathfrak{S}}^{11}- \delta {\mathfrak{S}}^{22}+
\Omega^2 \left(\Gamma^{11}-\Gamma^{22}\right)\,,\label{eqv1}\\
\omega^2\Gamma^{12}+\omega\Omega\left(\Gamma^{11}-\Gamma^{22}\right) &=&2\delta {\mathfrak{S}}^{12}+
2\Omega^2\Gamma^{12} \,.\label{eqv2}
\end{eqnarray}
In order to solve equations (\ref{eqv1}), (\ref{eqv2}), where the
virial integrals are given by (\ref{vir11}), (\ref{vir12}), we have to
compute the surface integrals $\delta\mathfrak S^{12}$ and
$\delta\mathfrak S^{11}-\delta\mathfrak S^{22}$.

In the case of perturbations of the form (\ref{tor}), they are (see
Appendix \ref{toroidal})
\begin{eqnarray}
\delta\mathfrak S^{12}&=& -\frac{T}{2}\beta\pi C_{D-2}\int_0^a
d\varpi\phi\varpi^4Q(\varpi)[f(\varpi)]^{D-3}\\
\delta\mathfrak S^{11}-\delta\mathfrak S^{22} &=&-T\alpha\pi C_{D-2}
\int_0^a d\varpi\phi\varpi^4Q(\varpi)
[f(\varpi)]^{D-3}\,.
\end{eqnarray}
where
\begin{equation}
Q=\frac{1}{\varpi ^2}\left (\frac{\phi'(\phi^2-2)}{(1+\phi^2)^{5/2}}
+\phi\frac{2+3\phi^2}{\varpi(1+\phi^2)^{3/2}}-\frac{D-3}{z}
\frac{\phi^2}{(1+\phi^2)^{3/2}} \right )\,. \label{QQ}
\end{equation}
Therefore equations (\ref{eqv1}), (\ref{eqv2}) are equivalent to
\begin{eqnarray}
\omega^2\alpha-2\omega\Omega\beta&=&-2{\mathfrak I}\alpha+
2\Omega^2\alpha\nonumber\\
\omega^2\beta+2\omega\Omega\alpha&=&-2{\mathfrak I}\beta+ 2\Omega^2 \beta\label{systv}
\end{eqnarray}
where
\begin{equation}
{\mathfrak{I}}\equiv\frac{T(D-2)}{4\rho}
\frac{\int_0^a\phi\varpi^4Q(\varpi) [f(\varpi)]^{D-3}d\varpi}{
\int_0^a\varpi^3\left[f(\varpi)\right]^{D-2} d\varpi}\,. \label{jota}
\end{equation}
Setting $\omega^2=-\sigma^2$, the algebraic system (\ref{systv}) admits solutions
\begin{equation}
\sigma=\Omega\pm\sqrt{2{\mathfrak I}-\Omega^2}\,.\label{freq}
\end{equation}
A neutral mode of oscillation ($\sigma=0$) occurs at $\Omega^2={\mathfrak I}$, but (in the absence of
dissipative mechanisms as stressed in \cite{chandra}) the drop is still stable in the vicinity of this point.
If
\begin{equation} \Omega^2\le2{\mathfrak I}\,, \label{instcond}\end{equation}
then $\sigma$ is real and the perturbation is stable; if the rotation frequency of the drop is larger, then the
drop is unstable against toroidal perturbations. To ascertain stability or instability, we thus have to use the
results of section \ref{sec:eqcasea} for the equilibrium figures of rotating drops and insert the numerically
generated shapes $z=z(\varpi)$ in (\ref{jota}) and (\ref{freq}).

\begin{figure}[ht]
\centerline{\includegraphics[width=8 cm,height=6 cm] {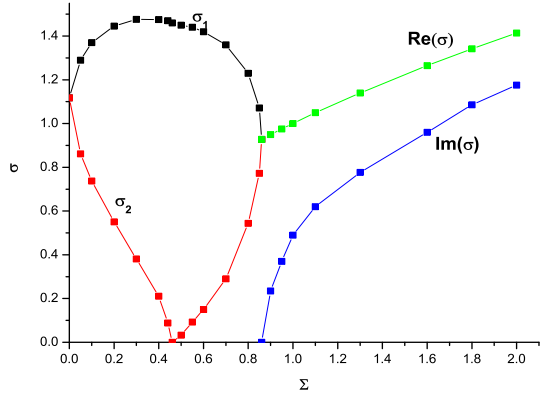}} \caption{The dependence on the quantity
$\Sigma$ of the characteristic frequencies of the toroidal modes of a rotating four-dimensional drop in $D=4$.
The diagram is very similar to the three dimensional one ($D=3$) presented in Figure 3 of Chandrasekhar
\cite{chandra}. As a check on our numerical code, we verify that in the non-rotating limit our numerical
results agree well with the analytical prediction $\sigma=\sqrt{D+1}/2$.} \label{fig:frediagram}
\end{figure}

If we measure $\varpi$ and $z$ in units of $a$ and $\Omega$ in units
of $\sqrt{\frac{8T}{\rho a^3}}$ we get
\begin{equation} \sigma=\sqrt{\Sigma_a} \pm\sqrt{2J-\Sigma_a}\,,\label{sigma} \end{equation}
where the dimensionless quantity $J$ is
\begin{equation}
J\equiv\frac{(D-2)}{32} \frac{\int_0^a\phi\varpi^4Q(\varpi)
[f(\varpi)]^{D-3}d\varpi}{
\int_0^a\varpi^3\left[f(\varpi)\right]^{D-2} d\varpi}\,.
\end{equation}
The results of solving (\ref{sigma}) are displayed in Fig. \ref{fig:frediagram} for $D=4$, which is a first
step towards reconstructing the whole bifurcation diagram. In Fig. \ref{fig:frediagram},  $\sigma_1$ is (for
real solutions) the largest of the two roots in (\ref{sigma}) , i.e., as long as $2J>\Sigma_a$. For
$2J<\Sigma_a$, we display only one root (the other root is the complex conjugate of this) by plotting the real
and imaginary components of $\sigma$.

In three dimensions ($D=3$) it was found by Chandrasekhar \cite{chandra} that rotating drops exhibit a neutral
mode of oscillation at $\Sigma=0.4587$. He also found that rotating drops become unstable for $\Sigma>0.84$. We
find that for general $D$ the results are similar. For instance, in $D=4$ there is a neutral mode of
oscillation at $\Sigma_a=0.46$, and above the critical $\Sigma_a=0.86$ the drops are unstable.
The shapes of the drops at these values of $\Sigma$ are displayed in
Figure \ref{fig:eqfigsa}. Results for higher $D$ are very similar. Our
numerical method only allowed to probe dimensions up to $D=6$, for
larger $D$ we loose accuracy and the results cannot be trusted.
\section{\label{sec:cylindersurface} Stability of fluid cylinders
held by surface tension: the Rayleigh-Plateau instability}
To end our discussion on fluids held by surface tension and also as a
warm up exercise for the up-coming Sections, we will review the
results of \cite{cardosodiasglrp}, describing a simple capillary
instability - the Rayleigh-Plateau instability
\cite{rayleigh,tomotika} - as a good model for the Gregory-Laflamme
instability.  Most calculations were omitted in \cite{cardosodiasglrp}
so we present the full detailed computations here.
\subsection{\label{subsec:rp} Rayleigh-Plateau instability in higher dimensions}
\begin{figure}[ht]
\centerline{\includegraphics[width=8 cm,height=6 cm] {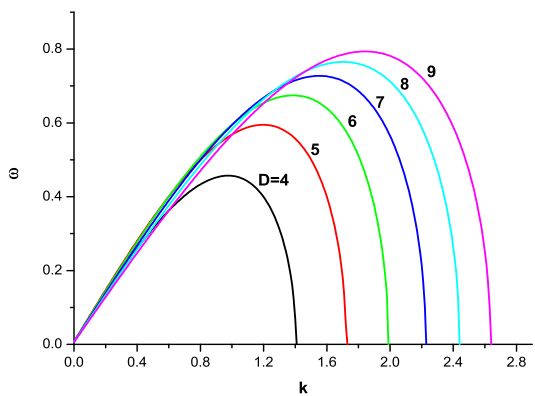}} \caption{The Rayleigh-Plateau instability of a
hyper-cylinder held together by surface tension, in several dimensions. Here the effective density were chosen
to match those of a higher-dimensional non-rotating black hole. The instability gets stronger as the spacetime
dimensionality increases, and the critical wavenumber (for which $\omega=0$) increases with $D$.}
\label{fig:insttimesurface}
\end{figure}

We take an infinite non-rotating cylinder of radius $R_0$ held
together by surface tension (therefore without
self-gravity). Neglecting the external pressure, the unperturbed
configuration is worked out easily to yield an internal pressure
$P_0=(D-2)T/R_0$, where $D$ is the total number of spatial
dimensions. We perturb slightly this configuration, according to
\begin{equation} r=R_0+\epsilon e^{\omega t+ikz}\,. \label{pertsurface} \end{equation}
The equations governing small perturbations are
\begin{eqnarray}
P_0+\delta P&=&T{\rm div}\vec{n}\,,\\
\frac{\partial \bf u}{\partial t}&=&-{\rm grad}
\Upsilon \label{gradupssurface}\,,\\
{\rm div} {\bf u}&=&0\,,\label{divergenlesssurface}\\
\Upsilon &\equiv& \frac{\delta P}{\rho}\label{defupssurface}\,, \end{eqnarray}
where $P_0$ is the pressure in the undisturbed cylinder and ${\bf u}$
is the fluid velocity. For instance, $u_r$ is the radial component of
the velocity vector. Equation (\ref{gradupssurface}) yields
\begin{eqnarray}
u_r&=&-\frac{1}{\omega}\frac{d}{dr}\Upsilon\,,\label{ursurface}\\
u_z&=&\frac{-ik}{\omega}\Upsilon\,.\label{uzsurface} \end{eqnarray}
Equation (\ref{divergenlesssurface}) implies that
\begin{equation} \frac{du_r}{dr}+\frac{D-2}{r}u_r+\frac{du_z}{dz}=0 \,. \end{equation}
Substituting in this the relations (\ref{ursurface}) and
(\ref{uzsurface}) we get
\begin{equation} \frac{d^2\Upsilon}{dr^2}+\frac{D-2}{r}\frac{d\Upsilon}{dr}-k^2
\Upsilon =0\,, \end{equation}
with (well-behaved at the origin) solution
\begin{equation} \Upsilon=C r^{3/2-D/2}I_{\alpha}(kr)\,, \label{Upsilonsurface} \end{equation}
where $\alpha=(D-3)/2$. By relation (\ref{ursurface}) we get
\begin{equation} u_r=-C\frac{r^{3/2-D/2}}{\omega}\left ( \frac{3-D}{2r}
I_{\alpha}(kr) +kI'_{\alpha}(kr)\right ) \,, \end{equation}
where $I'_{\alpha}(x)$ stands for the derivative of the Bessel
function with respect to the argument $x$. Now, for this velocity to
be compatible with the radial perturbation (\ref{pertsurface}) one
must clearly have
\begin{equation} \epsilon \,e^{\omega t+ikz}=-C\frac{R_0^{3/2-D/2}}{\omega^2} \left
( \frac{3-D}{2R_0} I_{\alpha}(kR_0)+kI'_{\alpha}(kR_0)\right )
\,.\label{epsilonsurface} \end{equation}
With the displacement (\ref{pertsurface}) we have to first order in $\epsilon$
\begin{equation} \frac{\delta P}{\rho}=-\frac{T}{R_0^2\rho}\left
(D-2-(kR_0)^2\right) \epsilon e^{\omega t+ikz}\,.\label{deltaP} \end{equation}
Using (\ref{Upsilonsurface}) and (\ref{epsilonsurface}) together with
(\ref{defupssurface}), (\ref{deltaP}) we finally find
\begin{equation} R_0^2\omega^2=\frac{T}{\rho}\frac{\left ( \frac{3-D}{2R_0}
I_{\alpha}(kR_0)+ kI'_{\alpha}(kR_0)\right )}{I_{\alpha}}\left (
D-2-(kR_0)^2 \right ) \,, \label{srefacerp}\end{equation}
where $I'_{\alpha}(x)$ stands for the derivative of the Bessel
function with respect to the argument $x$. This is the sought relation
yielding the characteristic frequencies of an infinite cylinder of
fluid held by surface tension. Formula (\ref{srefacerp}) constitutes a
generalization of a classic result by Rayleigh \cite{rayleigh}, and it
predicts that for $kR_0<\sqrt{D-2}$ the cylinder is unstable against
this kind of ``beaded'' perturbations. In Figure
\ref{fig:insttimesurface} we show $\omega$ as function of wavenumber
$k$ for several space dimensions $D$ (actually what is shown in the
plots are dimensionless quantities $\omega R_0$ and $kR_0$).
In Section \ref{sec:cylindergravity} we will discuss this instability
in more detail, when we compare it to its self-gravitating
counter-part.
\subsection{Rayleigh-Plateau and Gregory-Laflamme}
In General Relativity, there is an interesting class of black objects (objects with an event horizon) without
spherical topology, the black branes \cite{horowitzstrominger}. A broad class of these objects are unstable
against gravity, in a mechanism known as Gregory-Laflamme (GL) instability \cite{gl}. This is a beaded
instability that makes any small perturbation with wavelength $\lambda$ of the order of, or larger than, the
radius of the cylinder $R_0$ grow exponentially with time. It is a very robust long-wavelength instability. In
\cite{cardosodiasglrp} it was proposed to mimic the GL instability with the Rayleigh-Plateau instability, based
on the suggestive appearance of the second law of black hole mechanics which endows the horizon with an
effective surface tension. It was found that all aspects of the GL instability could be accounted for by the
Rayleigh-Plateau instability in fluid cylinders. In some sense, the effective surface tension can mimic all
gravitational aspects of black strings. If this is indeed correct, we should be able to predict the effects of
rotation and charge on the GL instability, as discussed in the following.
\subsubsection{\label{sec:extrapolationsurface}Effects of magnetic fields}
Even though not pursued here, a generalization for charged and/or rotating cylinders is of interest. In any
case, we can lean on the four-dimensional results ($D=3$ spatial dimensions), for which the charged and
rotating cases have been worked out \cite{chandrabookstability,interfacial}. A magnetic field has a stabilizing
effect \cite{chandrabookstability}, which is in agreement with the fully relativistic results for the
Gregory-Laflamme instability, where it has been shown that adding charge to the black strings stabilizes it
\cite{gl}. There is good evidence that extremal charged black strings are marginally stable, and indeed in the
capillary case (cylinder held by surface tension) it is possible to completely stabilize the cylinder if the
magnetic field has a certain strength \cite{chandrabookstability} (dependent on the conductivity of the fluid),
so this may simulate the marginal stability of extremal black strings very well \cite{gl}. Nevertheless an
extension of the calculations presented here to the case of magnetically charged cylinders with surface tension
is needed.

\subsubsection{\label{sec:extrapolationsurface2} Effects of rotation}
The effects of rotation on the capillary Rayleigh-Plateau instability can be discussed more intuitively. The
centrifugal force, scaling with $r$, contributes with a de-stabilizing effect since there is an increased
pressure under a crest but a reduced pressure under a trough. So in principle, the threshold wavelength should
be smaller \cite{interfacial}, i.e., the instability should get stronger when rotation is added. It is natural
to expect that rotation will lift the degeneracy of the modes and that unstable modes will in general be non
axi-symmetric. What can we expect for the general relativistic GL instability once rotation is added? A very
naive reasoning is that since extremal charged black strings are marginally stable then perhaps the same
happens with rotation, and so rotation has a stabilizing effect. This is incorrect as a simple argument shows.
Consider a rotating black string in $D+1$ spacetime dimensions (with again $D$ being the number of spatial
dimensions) of the form
\begin{equation} g={\rm Kerr}^{D} \times {\bf R}\,,\end{equation}
where ${\rm Kerr}^{D}$ is the metric of higher dimensional Myers-Perry rotating black hole in $D$ spacetime
dimensions, with only one angular momentum parameter (case I). If we focus on $D>4$ the angular momentum of
these black holes is unbounded and one can show \cite{emparanmyers} that for very large angular momentum
\begin{equation} {\rm Kerr}^D \sim S_{chw}^{D-2}\times {\bf R^2}\,,{\rm for \,\,large\,\,rotation} \end{equation}
where $S_{chw}^{D-2}$ is the metric of a higher dimensional Schwarzschild black hole, often referred to as
Tangherlini \cite{tangherlini} black holes. We thus finally get
\begin{equation} {\bf g}=S_{chw}^{D-2} \times {\bf R^3}\,,\end{equation}
This is the metric of a {\it non-rotating} black membrane in $D+1$ spacetime dimensions extended along the $3$
dimensions. But this is precisely the geometry considered by Gregory and Laflamme \cite{gl}, and so it is
unstable. While this argument says little about the strength of the instability as function of rotation, it
does predict that rotating branes are still unstable. Putting this and the fluid membrane argument together we
predict that rotation makes the GL instability stronger. One can also use an entropy argument, similar to the
one in \cite{mu-inpark}, to show that rotating black strings should be unstable.
 Again, an extension of the calculations presented in
Section \ref{subsec:rp} to the case of rotating cylinders with surface tension is needed. The $D=3$ case,
studied in \cite{surfacetensionrotating}, indicates that other effects may come into play with rotation. For
instance, other instabilities seem to manifest themselves for high enough rotation, so this is a particularly
exciting topic to consider.

\part{Fluids held by (Newtonian) self-gravity}

\vskip 1cm

\section{\label{sec:cylindergravity}Stability of self-gravitating fluid
cylinders: the Dyson-Chandrasekhar-Fermi instability}
It is only appropriate that we start the study of self-gravitating
fluids by examining the (self-gravitating) counterpart of the cylinder
studied in the last Section \ref{sec:cylindersurface}.

A simple consistent solution can be found in $D$-dimensions, which corresponds to a static self-gravitating
infinite cylinder made of fluid with a density $\rho$. For the moment we will not consider rotational effects
(these will be discussed later), and we will focus on incompressible fluids. The stability of such a solution
for the case of 3 spatial dimensions was studied for the first time by Chandrasekhar and Fermi
\cite{chandrafermi} (see also \cite{chandrabookstability}), who showed that such a solution is unstable against
cylindrically symmetric perturbations (those for which the cross-section is still a circle, with a
$z-$dependent radius), and stable against all others. If the radius of the cylinder is $R_0$, this ``varicose''
instability (in Chandrasekhar's \cite{chandrabookstability} terminology) sets in for wavelengths $\lambda>2\pi
R/1.0668$.

A similar phenomenon had however been studied already in 1892 by Dyson \cite{dyson}, who investigated the
stability of self-gravitating rings of fluid. In his terminology, the ring is stable against fluted (the ring
remains symmetrical about its axis but the cross-section is deformed) and twisted (the cross-section remains
circular but the circular axis of the ring is deformed) perturbations. He also showed that sufficiently thin
rings are unstable against a third kind of perturbation, so-called beaded perturbations (those for which the
circular axis is undisturbed, but the cross-section is a circle of variable radius). But these correspond
precisely to Chandrasekhar-Fermi type of perturbation (with the added bonus that Dyson effectively shows that
the instability goes away if the length of the cylinder or ring is less than the critical wavelength), and thus
we shall refer to this classical fluid instability as the Dyson-Chandrasekhar-Fermi (DCF) instability.

\subsection{A self-gravitating, non-rotating $D$-dimensional fluid cylinder}
We will now generalize the DCF instability for an arbitrary number of dimensions. Consider an infinite cylinder
of incompressible fluid, with density $\rho$ and radius $R$. To fix conventions we consider spacetime with $1$
(time)$+1$ (axis of cylinder)$+D-1=D+1$ total number of spacetime dimensions. In the following the coordinates
$r\,,z$ have their usual meaning in polar coordinates. The gravitational potential ${\cal B}({\bf x})$ inside
and exterior to the cylinder satisfies respectively
\begin{eqnarray}
\nabla^2{\cal B}({\bf x})^i&=&-(D-2)C_D G \rho\,,\label{vi}\\
\nabla^2{\cal B}({\bf x})^e&=&0\,. \label{ve} \end{eqnarray}
Now, in $D$ spatial dimensions with cylindrical symmetry we have
\begin{equation} \nabla^2=\frac{1}{r^{D-2}}\frac{d}{dr}\left (r^{D-2}\frac{d}{dr} \right
)+\frac{d^2}{dz^2}\,.\label{laplacian} \end{equation}
Since we consider a cylinder with constant radius (along the axis in the $z$-direction), the last term drops
off and we get the following solution (we demand that both the potential and its derivative be continuous
throughout)
\begin{eqnarray}
{\cal B}({\bf x})^i_0&=& -\frac{(D-2)C_D G\rho}{2(D-1)}r^2\,,\label{Vi}\\
{\cal B}({\bf x})^e_0&=&-\frac{(D-2)C_D}{2} G \rho R_0^2 \log r/R_0-
\frac{D-2}{4}C_D G \rho R_0^2\,, \quad
D=3\,.\\
{\cal B}({\bf x})^e_0&=&-\frac{(D-2)C_D G \rho R_0^2}{2(D-1)}
\left ( \frac{D-1}{(D-3)}-
\frac{2R_0^{D-3}}{(D-3)r^{D-3}}\right )\,, \quad D>3\,. \end{eqnarray}
Hydrostatic equilibrium demands that the pressure $P_0$ satisfies
\begin{equation} \frac{d}{dr}\left (\frac{P_0}{\rho}-{\cal B}({\bf x})\right )=0 \,, \end{equation}
which yields, upon requiring $P_0(r=R_0)=0$,
\begin{equation} P_0=-\frac{(D-2)C_D G_L \rho ^2}{2(D-1)}(r^2-R_0^2)\,. \end{equation}
%
\subsection{Axisymmetric perturbations}
\begin{figure}[ht]
\centerline{\includegraphics[width=8 cm,height=3 cm] {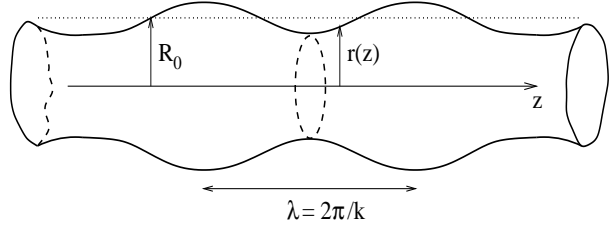}} \caption{Black strings and fluid cylinders
are unstable to perturbations on the extended dimension, i.e., along the axis of the cylinder. Ripples
propagating along this axis grow exponentially with time for wavelengths of order of the radius of the
cylinder.} \label{fig:inst}
\end{figure}
Suppose now the system is subjected to a perturbation (see Fig. \ref{fig:inst}) such that the boundary is
described by
\begin{equation} r=R_0+\epsilon e^{\omega t+ikz}\,. \label{pert} \end{equation}
The equations governing small perturbations are
\begin{eqnarray}
\nabla^2 \delta {\cal B}({\bf x})&=&0\,,\\
\frac{\partial \bf u}{\partial t}&=&-{\rm grad} \Upsilon \label{gradups}\,,\\
{\rm div} {\bf u}&=&0\,,\label{divergenless}\\
\Upsilon &\equiv& \frac{\delta P}{\rho}-\delta {\cal B}({\bf x})^i
\label{defups}\,. \end{eqnarray}
Here ${\bf u}$ stands for the velocity of the perturbed fluid. Let's solve for the first of these equations.
When solving for the gravitational potential, the last term in (\ref{laplacian}) cannot be neglected. We find
\begin{eqnarray}
{\cal B}({\bf x})^e&=&{\cal B}({\bf x})^e_0+A\epsilon e^{\omega t+ikz}
r^{3/2-D/2}K_{\alpha}(kr) \,,\\
{\cal B}({\bf x})^i&=&{\cal B}({\bf x})^i_0+B\epsilon e^{\omega t+ikz}
r^{3/2-D/2}I_{\alpha}(kr)\,, \end{eqnarray}
where
\begin{equation} \alpha=\frac{D-3}{2}\,, \end{equation}
and $K_{\alpha}$ and $I_{\alpha}$ are Bessel functions of argument
$\alpha$. The constants $A$ and $B$ are determined by requiring
continuity of the potential and its derivative on the boundary. By
keeping only dominant terms in $\epsilon$ we find
\begin{eqnarray}
A&=&(D-2)C_D G \rho R_0 I_{\alpha}(kR_0)\,,\\
B&=&(D-2)C_D G \rho R_0 K_{\alpha}(kR_0)\,. \end{eqnarray}
Therefore we have
\begin{equation} \delta {\cal B}({\bf x})^i=\epsilon (D-2)C_D G \rho R_0
  K_{\alpha}(kR_0) e^{\omega t+ikz}r^{3/2-D/2}I_{\alpha}(kr)\,.  \end{equation}
We also have $u \sim e^{i\omega t+ikz}$, therefore Eq. (\ref{gradups}) yields
\begin{eqnarray}
u_r&=&-\frac{1}{\omega}\frac{d}{dr}\Upsilon\,,\label{ur}\\
u_z&=&\frac{-ik}{\omega}\Upsilon\,.\label{uz} \end{eqnarray}
(Here $u_r$, for instance, is the radial component of the velocity
vector). Equation (\ref{divergenless}) implies that
\begin{equation} \frac{du_r}{dr}+\frac{D-2}{r}u_r+\frac{du_z}{dz}=0 \,. \end{equation}
Substituting the relations (\ref{ur}) and (\ref{uz}) we get
\begin{equation}
\frac{d^2\Upsilon}{dr^2}+\frac{D-2}{r}\frac{d\Upsilon}{dr}-k^2\Upsilon
=0\,,
\end{equation}
with (well-behaved at the origin) solution
\begin{equation} \Upsilon=C r^{3/2-D/2}I_{\alpha}(kr)\,. \label{Upsilon} \end{equation}
By relation (\ref{ur}) we get
\begin{equation} u_r=-C\frac{r^{3/2-D/2}}{\omega}\left ( \frac{3-D}{2r}
I_{\alpha}(kr) +kI'_{\alpha}(kr)\right ) \,, \end{equation}
where $I'_{\alpha}(x)$ stands for the derivative of the Bessel
function with respect to the argument $x$. Now, for this velocity to
be compatible with the radial perturbation (\ref{pert}) one must
clearly have
\begin{equation} \epsilon \,e^{\omega t+ikz}=-C\frac{R_0^{3/2-D/2}}{\omega^2}
\left ( \frac{3-D}{2R_0}
I_{\alpha}(kR_0)+kI'_{\alpha}(kR_0)\right ) \,.\label{epsilon} \end{equation}
Our problem is completely specified once we impose the boundary
condition that the pressure must vanish at the boundary,
\begin{equation} \delta P+\epsilon \frac{\partial P_0}{\partial r}=0\,,
\Rightarrow \rho \Upsilon+\rho \delta {\cal B}({\bf
x})^i+\epsilon \frac{\partial P_0}{\partial r}=0\,. \end{equation}
This gives us finally
\begin{equation} \frac{\omega^2}{(D-2)C_D G\rho}=\frac{(3-D)I_{\alpha}(kR_0)+
kR_0I'_{\alpha} (kR_0)}{2I_{\alpha}(kR_0)}\left
(I_{\alpha}(kR_0)K_{\alpha}(kR_0)-\frac{1}{D-1}\right )\,.
\label{dcfinstability} \end{equation}
\begin{figure}[ht]
\centerline{\includegraphics[width=8 cm,height=6 cm] {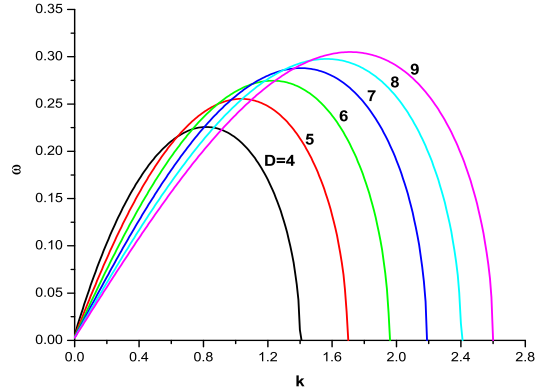}} \caption{The DCF instability of a
self-gravitating hyper-cylinder in several dimensions. Here the effective densities were chosen to match those
of a higher-dimensional non-rotating black hole. The instability gets stronger as the spacetime dimensionality
increases, and the critical wavenumber (for which $\omega=0$) increases with $D$.} \label{fig:insttime}
\end{figure}
In Fig. \ref{fig:insttime} we plot the dispersion relation $\omega(k)$ for several values of the dimensionality
$D$, using Eq. (\ref{dcfinstability}). When $D$ increases the maximum $\omega$ grows which means that the
instability gets stronger with space dimensionality. The threshold wavenumber $k_c$, defined as the wavenumber
for which $\omega=0$, also increases with $D$. In Table \ref{tab:thresholdmode} we present the threshold
wavenumber for selected values of $D$.
\begin{table*}
\caption{\label{tab:thresholdmode} Dimensionless threshold wavenumber
$k R_0$ for the DCF instability of a higher dimensional fluid
cylinder, and the corresponding threshold wavenumber for
Rayleigh-Plateau and Gregory-Laflamme instability (these last data are
taken from \cite{kolsorkin}). The cylinder has $D-1$ dimensions with
radius $R_0$ plus an extended direction along $z$. For $D=3$, not
shown in the Table, we get $kR_0=1.06672$, in complete agreement with
Chandrasekhar and Fermi's \cite{chandrafermi} result.}
\begin{ruledtabular}
\begin{tabular}{ccccccccc}  \hline
$D$ &4&5&6&7&8&9&49&99\\ \hline
{\rm DCF}&            1.41      &1.70  &1.96 &2.20
&2.41&2.61&6.85&9.85\\
%
{\rm Rayleigh-Plateau}&1.41     &1.73  &2.00 &2.24
&2.45&2.66&6.78&9.80\\
%
{\rm Gregory-Laflamme}&0.876 &1.27 &1.58  &1.85 &2.09&2.30&6.72&9.75\\
%
\end{tabular}
\end{ruledtabular}
\vskip -2mm
\end{table*}

For large $D$ we can obtain an analytical expression for the threshold
wavenumber. In fact we can expand the Bessel functions as
\cite{stegun}:
\begin{eqnarray} I_{\alpha}(kR_0)&\sim& (kR_0/2)^{\alpha}\left (\frac{1}{\alpha!}+
\frac{(kR_0)^2}{4(\alpha+1)!}\right )\,,\\
 K_{\alpha}(kR_0)&\sim& \frac{1}{2}(kR_0/2)^{-\alpha}\left (
(\alpha-1)!-\frac{(\alpha-2)!(kR_0)^2}{4}\right )\,. \end{eqnarray}
We get
\begin{equation} K_{\alpha}(kR_0)I_{\alpha}(kR_0)-\frac{1}{D-1} \sim
\frac{2}{D^2}-\frac{(kR_0)^2}{4\alpha^3}\,\Rightarrow\,\, kR_0 \sim
\sqrt{D}\,. \end{equation}
A comparison between these results and the ones in Section \ref{sec:cylindersurface} is instructive. The Figs.
\ref{fig:insttimesurface} and \ref{fig:insttime} show a remarkably similar behavior for both these
instabilities. The threshold wavenumber (for which $\omega=0$), shown in Table \ref{tab:thresholdmode} is also
very similar to the corresponding one in both the Gregory-Laflamme and the Rayleigh-Plateau instability.
Finally the $D \rightarrow \infty$ limit of the three instabilities is the same, $kR_0 \sim \sqrt{D}$, in the
sense that the threshold wavenumber coincides in this limit.
\subsection{DCF and Gregory-Laflamme}
In \cite{cardosodiasglrp}, the authors proposed to mimic the GL instability with the Rayleigh-Plateau
instability, based on the suggestive appearance of the second law of black hole mechanics which endows the
horizon with an effective surface tension. What we now propose is that, while the fluid membrane analogy is
extremely powerful (and perhaps more trustworthy when it comes to mimic black objects), the Newtonian classical
counterpart of the GL instability is the DCF instability. The GL instability should be the limit of the DCF
instability when the density is so large that the cylinder is in fact a black string. Not only are the details
of the DCF instability very similar to the GL instability (for instance, compare Figure \ref{fig:insttime} with
Figure 1 of \cite{gl}) but for large $D$ we get exact agreement for the threshold wavenumber. In Table
\ref{tab:thresholdmode} we compare the threshold wavenumber for the three different types of instabilities
(DCF, Rayleigh-Plateau and GL). The agreement is indeed very good, thus providing a convincing evidence for our
conjectured analogy between DCF/Rayleigh-Plateau instabilities for fluids and GL instability for black strings.

It has been shown by Chandrasekhar and Fermi \cite{chandrafermi} (see also Chandrasekhar
\cite{chandrabookstability}) that the DCF instability is rooted in the sign change of potential energy for
these varicose or beaded perturbations. But this is precisely what happens for the capillary of
Rayleigh-Plateau instability (see Section \ref{sec:cylindersurface} and discussion in \cite{cardosodiasglrp}),
which explains why both of them yield so similar results. This is another pleasant indication that these
instabilities are related. Based on this agreement we can discuss what to expect for rotating and charged
cylinders and discuss the significance of the results for the GL instability.
\subsection{\label{sec:extrapolation}Effects of rotation and magnetic fields}
Even though not pursued here, a generalization for charged and/or rotating cylinders is of great interest. In
any case, we can lean on the four-dimensional results ($D=3$ spatial dimensions), for which the charged
\cite{chandrabookstability,chandrafermi,shivamoggi} and rotating cases \cite{karnik,robe,luyten} have been
worked out.

It has been shown that a magnetic field has a {\it stabilizing} effect, reducing the threshold wavenumber,
i.e., increasing the minimum wavelength at which instability occurs. This is in agreement with the fully
relativistic results for the Gregory-Laflamme instability, where it has been shown that adding charge to the
black string stabilizes it \cite{gl}. There is good evidence that the extremal case is marginally stable, and
this could be perhaps the only real signature of an event horizon \footnote{The fact that extremal charged
strings are marginally stable may not be directly related to horizons in themselves but rather to pure general
relativistic effects. Thorne \cite{thornemelvin} has shown for instance that the Melvin universe in general
relativity is absolutely stable. The Newtonian counterpart of the Melvin universe, on the other hand, is
unstable \cite{wheelermelvin}.}, because in the horizonless, Newtonian case, ``no magnetic field, however
strong, can stabilize the cylinder for disturbances of all wavelengths: the gravitational instability of long
waves will persist'' \cite{chandrabookstability}. As we discussed in Section \ref{sec:cylindersurface}, a
capillary model may be more accurate to describe extremal objects.

Focusing now on the effect of rotation, it has been shown \cite{karnik} (in three spatial dimensions) that
rotation has a {\it destabilizing} effect. This agrees with the phenomenological arguments in
\cite{cardosodiasglrp} for cylinders held by surface tension (see also \cite{interfacial}) of the capillary
model. It also agrees with the arguments in Section \ref{sec:cylindersurface} suggesting that rotating black
strings are unstable and thus a solid conjecture is that the Gregory-Laflamme instability of black strings gets
stronger if one adds rotation. Stronger here means the threshold wavenumber decreases (the threshold wavelength
increases). An interesting extension of the results presented here would be to consider a $D$-dimensional
rotating cylinder of fluid.
\section{\label{sec:torusselfgravitating} Self-gravitating torus and black rings}
Toroidal shapes are thought to be common configurations in the
universe. In some of them these rings or tori are associated with a
massive central object, such as Saturn's rings, whereas in others the
central object plays almost no role at all (see \cite{wong} and
references therein). Toroidal configurations in full general
relativity are also important (see \cite{ansorgringblackhole}), and
they sometimes appear in the form of temporary black toroids, having
an event horizon, in simulations of collapse of rotating matter
\cite{shapiro}.

The problem of a self-gravitating torus in Newtonian theory was
investigated by many authors, the first studies belonging to
Poincar\'e, Kowalewsky \cite{history} and Dyson \cite{dyson}. Dyson
\cite{dyson} considers a self-gravitating ring of radius $R$ and
thickness $2r_+$ (i.e. the cross-section of the ring is a circle with
radius $r_+$). For small $r_+/R$ he finds a power series solution (the
leading term had already been derived by Poincar\'e and Kowalewsky
\cite{history}) to the equilibrium problem as
\begin{equation} \frac{\Omega^2}{\pi G_L\rho}= \frac{r_+^2}{R^2}\left ( \log\left
(\frac{8R}{r_+}\right )-\frac{5}{4} \right
)-\frac{r_+^4}{8R^4}\left (\log\left (\frac{8R}{r_+} \right )-
\frac{5}{12}\right )+{\cal O}(r_+/R)^5 \,,\quad D=3\,.
\end{equation}
To generalize this result to arbitrary dimensions we first observe that the gravitational potential of a very
large ring of radius $R$ can be obtained by using as gaussian surface a torus. We get, for $D>3$,

\begin{equation} {\cal B}({\bf x}) \sim -\frac{(D-2)C_DGM}{4\pi (D-3)C_{D-2}R
r_+^{D-3}}\,, \quad D>3\,,\label{Vring} \end{equation}
where we expressed everything in terms of the physically more
significant quantity $M$, the total mass of the ring.
For rings with a large radius $R$ the hydrostatic equilibrium
condition can be written as \cite{dyson,chandrabookellipsoidal}
$\Omega^2R^2 \sim {\cal B}({\bf x})$. This gives us
\begin{equation} \Omega^2R^2 \sim \frac{(D-2)C_D M}{4\pi(D-3)C_{D-2}R \,r_+^{D-3}}
\label{equilibriumrings} \end{equation}
It seems plausible that this formula also describes general relativistic objects, as long as one uses an
``equation of state'' to deduce $r_+$ in terms of the other quantities. This is because we are considering
large rings, without really appealing to the internal structure of the object. For large rings, the dominating
force determining equilibrium should be well-described by Newtonian forces, even though general relativistic
effects become important ``locally'' and therefore it is the latter that determine $r_+(R,M)$. To plug-in the
information that we are dealing with a black object, we use the relation between $r_+$, $M$ and $R$. For a
black ring, the mass per length should be given, to a good accuracy, by the mass of a $D-1$ Tangherlini black
hole \cite{tangherlini}:
\begin{equation} (D-2)C_DG M \sim
4\pi(D-2)C_{D-2}R\,r_+^{D-3}\,. \label{massblackobjects} \end{equation}
Using (\ref{equilibriumrings}) and re-expressing everything in terms
of the angular momentum $J\sim MR^2\Omega$ we get
\begin{equation}
\frac{J^2}{M^2R^2}\sim 1\,. \label{newtring}
\end{equation}
We can compare this prediction against the exact general relativistic black ring solution \cite{Emparan:2001wn}
in five spacetime dimensions ($D=4$). For large black rings, Eq. (\ref{grring}),
$\frac{J}{MR}=\frac{\sqrt{2}}{3}$, in very good agreement with the Newtonian prediction (\ref{newtring}).
Moreover these arguments also lead to conjecture the existence of black rings in higher dimensions. Again the
case $D=3$ seems special for there are Dyson rings in this case (actually that's the case where they were
found), albeit with a peculiar logarithmic term. As we remarked at the end of Section \ref{sec:toroidalsurface}
there are no black rings in $D=3$ so this is an instance where the analogy breaks down. To end this section, we
note that another accurate and simple model was put forward by Hovdebo and Myers \cite{Hovdebo:2006jy} to
describe black rings in quasi-Newtonian terms, starting from the black string solution. Their conclusions are
basically the same as ours, with the exception that instead of working with Newtonian forces, they extremize
the entropy of the solution.

\section{\label{sec:MacLaurin} Self-gravitating, rotating spheroids: the MacLaurin sequence}

To complete our discussion of self-gravitating objects, we consider the generalization to $D$-spatial
dimensions of the simplest rotating objects: the MacLaurin sequence. Here we restrict ourselves to case of one
non-vanishing angular momentum (case $I$ in the notation of Section \ref{sec:symm}), corresponding to rotation
in the plane $x^1-x^2$; the coordinates are then decomposed as (\ref{coordinatesa})
\begin{equation}
x^i=(x^a,x^A)~;~~~a=1,2~;~~A=3,\dots,D\,.
\end{equation}
We will follow the approach of Chandrasekhar \cite{chandrabookellipsoidal}, based on the formalism of virial
tensors. The special case in which the rotation is zero (i.e., a self-gravitating globe), is considered in
Appendix \ref{app:gravitysphere}, where we compute the oscillation modes of a $D$-dimensional incompressible
fluid sphere.

\subsection{The MacLaurin sequence in three dimensions}
Here we briefly review the classical calculation of the MacLaurin
sequence due to Chandrasekhar. We refer to
\cite{chandrabookellipsoidal} for further details.

Let us consider a matter distribution, inside the volume $V$, with
density $\rho(\vec x)$. Its gravitational potential is
\begin{equation}
{\cal B}(\vec x)\equiv G\int_V\frac{\rho(\vec x')}{|\vec x-\vec
  x'|}d^3x'\,.
\end{equation}
We define the tensor
\begin{equation}
{\cal B}_{ij}(\vec x)\equiv G\int_V\rho(\vec x')\frac{(x^i-x^{\prime
    i})(x^j-x^{\prime j}) }{|\vec x-\vec
x'|^3}d^3x'\,, \end{equation}
which satisfies ${\cal B}_{ii}={\cal B}$. We also consider the
potential energy
\begin{equation} I\!{\cal B}\equiv-\frac{1}{2}\int_V\rho(\vec x){\cal B}(\vec x)d^3x\,, \end{equation}
and the tensor
\begin{equation}
I\!{\cal B}_{ij}\equiv-\frac{1}{2}\int_V\rho(\vec x){\cal B}_{ij}(\vec
x)d^3x\,.
\end{equation}
The following relation holds:
\begin{equation}
I\!{\cal B}_{ij}=\int_V\rho\frac{\partial{\cal B}}{\partial
  x^i}x^jd^3x\,.\label{relxpto}
\end{equation}
Indeed,
\begin{eqnarray}
&&\int_V\rho\frac{\partial{\cal B}}{\partial x^i}x^jd^3x=
-G\int_Vd^3x\int_Vd^3x'\rho(\vec x)\rho(\vec x') \frac{x^i-x^{\prime
i}}{|\vec x-\vec x'|^3}x^j= G\int_Vd^3x'\int_Vd^3x\rho(\vec
x')\rho(\vec x) \frac{x^i-x^{\prime i}}{|\vec x-\vec x'|^3}x^{\prime
j}\nonumber\\ &&=-\frac{1}{2}G\int_Vd^3x\int_Vd^3x'\rho(\vec x)\rho(\vec x')
\frac{(x^i-x^{\prime i})(x^j-x^{\prime j})}{|\vec x-\vec
x'|^3}\,.
\end{eqnarray}
A consequence of this formula is that $I\!{\cal B}_{ij}$ is symmetric
in its indices.

We then have the equations of motion
\begin{eqnarray} \ddot x^a&=&-\frac{\partial P}{\partial x^a}+\Omega^2\rho x^a+2
\rho\Omega\epsilon^{ab}\dot
x^b+\rho\frac{\partial {\cal B}}{\partial x^a}\nonumber\,,\\
\ddot z&=&-\frac{\partial P}{\partial z}+\rho\frac{\partial{\cal B}}{\partial z} \,,\end{eqnarray}
where $z=x^A=x^3$.

At equilibrium,
\begin{eqnarray} \Omega^2\rho x^a
-\frac{\partial P}{\partial x^a}+\rho\frac{\partial {\cal B}}{\partial x^a}&=&0\nonumber\,,\\
-\frac{\partial P}{\partial z}+\rho\frac{\partial{\cal B}}{\partial z}&=&0\,.\label{eqeq} \end{eqnarray}
We compute the virial integrals by multiplying equations (\ref{eqeq})
by $x^j$ and integrating in the volume domain $V$. The resulting
integrals are:
\begin{eqnarray}
\Omega^2\int_Vd^3x\rho x^ax^j&=&\Omega^2I_{aj}\nonumber\,,\\
-\int_V d^3x\frac{\partial P}{\partial x^i}x^j&=&\delta^{ij}\int_V d^3xP
\equiv\delta_{ij}
\Pi\nonumber\,,\\
\int_Vd^3x\rho\frac{\partial{\cal B}}{\partial x^i}x^j&=&I\!{\cal B}_{ij}\,. \end{eqnarray}
Thus we get virial equations
\begin{eqnarray}
\Omega^2I_{ab}+I\!{\cal B}_{ab}&=&-\Pi\delta_{ab}\nonumber\,,\\
\Omega^2I_{a3}+I\!{\cal B}_{a3}&=&0\nonumber\,,\\
I\!{\cal B}_{3a}&=&0~~~~~\hbox{(therefore $I\!{\cal B}_{a3}=I_{a3}=0$)}\nonumber\,,\\
I\!{\cal B}_{33}&=&-\Pi\,. \end{eqnarray}
The inertia tensor at equilibrium is then, in general,
\begin{equation}
I_{ij}=\left(\begin{array}{ccc} I_{11} & I_{12} & 0 \\
I_{21} & I_{22} & 0 \\ 0 & 0 & I_{33} \\ \end{array} \right)\,. \end{equation}
We note that equilibrium does not require necessarily $I_{11}=I_{22}$.

From the virial equations we get
\begin{equation}
I\!{\cal B}_{11}+\Omega^2I_{11}=I\!{\cal B}_{22}+\Omega^2I_{22}=
I\!{\cal B}_{33}\label{vir2}\,.
\end{equation}
There are two possible solutions of equations (\ref{vir2}). The first
is the simplest one, with
\begin{equation} I_{11}=I_{22}\,,~~~~~I\!{\cal B}_{11}=I\!{\cal B}_{22}\,, \end{equation}
namely, with axial symmetry; the rotation frequency is given by
\begin{equation} \Omega^2=\frac{I\!{\cal B}_{33}-I\!{\cal B}_{11}}{I_{11}}\,. \end{equation}
and can be easily computed; this solution corresponds to the MacLaurin sequence of oblate axially symmetric
ellipsoids. The second, more complicate, has $I_{11}\neq I_{22}$, and corresponds to the Jacobi sequence of
tri-axial ellipsoids; this solution exists only if the total angular momentum exceeds a critical value.
\subsection{The MacLaurin sequence in higher dimensions}
The generalization of the three dimensional formalism to higher dimensions is more or less straighforward.
 Let us consider a matter distribution in $D$ dimensions, inside the volume $V$, with
density $\rho(\vec x)$. Again, we define the gravitational potential
\begin{equation} {\cal B}(\vec x)\equiv G\int_V\frac{\rho(\vec x')}{|\vec x-\vec
  x'|^{D-2}}
d^Dx'\,. \label{gravd}
\end{equation}
and
\begin{equation} {\cal B}_{ij}(\vec x)\equiv\int_V\rho(\vec x')\frac{(x^i-x^{\prime i}) (x^j-x^{\prime j}) }{|\vec x-\vec
x'|^D}d^Dx'\,, \end{equation}
which satisfies ${\cal B}_{ii}={\cal B}$. We also consider the
potential energy
\begin{equation} I\!{\cal B}\equiv-\frac{1}{2}\int_V\rho(\vec x){\cal B}(\vec x)d^Dx\,. \end{equation}
and the tensor
\begin{equation}
I\!{\cal B}_{ij}\equiv-\frac{1}{2}\int_V\rho(\vec x){\cal B}_{ij}(\vec
x)d^Dx\,.
\end{equation}
We can check that the following relation, which is the extension of (\ref{relxpto}), holds:
\begin{equation}
I\!{\cal B}_{ij}=\frac{1}{D-2}\int_V\rho\frac{\partial{\cal
    B}}{\partial x^i}x^jd^Dx\,.
\end{equation}
Indeed,
\begin{eqnarray}
&&\int_V\rho\frac{\partial{\cal B}}{\partial x^i}x^jd^Dx=
-(D-2)G\int_Vd^Dx\int_Vd^Dx'\rho(\vec x)\rho(\vec x')
\frac{x^i-x^{\prime i}}{|\vec x-\vec x'|^D}x^j=
(D-2)G\int_Vd^Dx'\int_Vd^Dx\rho(\vec x')\rho(\vec x)
\frac{x^i-x^{\prime i}}{|\vec x-\vec x'|^D}x^{\prime j}\nonumber\\
&&=-\frac{D-2}{2}G\int_Vd^Dx\int_Vd^Dx'\rho(\vec x)\rho(\vec x')
\frac{(x^i-x^{\prime i})(x^j-x^{\prime j})}{|\vec x-\vec x'|^D}\,.
\end{eqnarray}
Thus $I\!{\cal B}_{ij}$ is symmetric in its indices, as it was in three dimensions.

The equations of motion yield
\begin{eqnarray}
\ddot x^a&=&-\frac{\partial P}{\partial x^a}+\Omega^2\rho
x^a+2\rho\Omega\epsilon^{ab}
\dot x^b+\rho\frac{\partial {\cal B}}{\partial x^a}\nonumber\,,\\
\ddot x^A&=&-\frac{\partial P}{\partial x^A}+\rho\frac{\partial{\cal B}}{\partial x^A}\,.
\end{eqnarray}
At equilibrium we get,
\begin{eqnarray}
\Omega^2\rho x^a
-\frac{\partial P}{\partial x^a}+\rho\frac{\partial {\cal B}}{\partial x^a}&=&0\nonumber\,,\\
-\frac{\partial P}{\partial x^A}+\rho\frac{\partial{\cal B}}{\partial
  x^A}&=&0\,.\label{eqeqd}
\end{eqnarray}
We compute the virial integrals by multiplying equations (\ref{eqeqd})
by $x^j$ and integrating in the volume domain $V$. The resulting
integrals are:
\begin{eqnarray}
\Omega^2\int_Vd^Dx\rho x^ax^j&=&\Omega^2I_{aj}\nonumber\,,\\
-\int_V d^Dx\frac{\partial P}{\partial x^i}x^j&=&\delta^{ij}\int_V d^DxP\equiv\delta_{ij}
\Pi\nonumber\,,\\
\int_Vd^Dx\rho\frac{\partial{\cal B}}{\partial x^i}x^j&=&(D-2)I\!{\cal B}_{ij}\,. \end{eqnarray}
Thus we get the virial equations
\begin{eqnarray}
\Omega^2I_{ab}+(D-2)I\!{\cal B}_{ab}&=&-\Pi\delta_{ab}\nonumber\,,\\
\Omega^2I_{aA}+(D-2)I\!{\cal B}_{aA}&=&0\nonumber\,,\\
(D-2)I\!{\cal B}_{Aa}&=&0
~~~~~\hbox{(therefore $I\!{\cal B}_{aA}=I_{aA}=0$)}\nonumber\,,\\
(D-2)I\!{\cal B}_{AB}&=&-\Pi\delta_{AB}\,. \end{eqnarray}
The inertia tensor at equilibrium is then, in general,
\begin{equation}
I_{ij}=\left(\begin{array}{cccc} I_{11} & I_{12} & 0 &  \\
I_{21} & I_{22} & 0 &  \\ 0 & 0 & I_{33} &  \\
 &   &  & \ddots \\ \end{array} \right)\,.
\end{equation}
We note that equilibrium does not require necessarily $I_{11}=I_{22}$.

From the virial equations we get
\begin{equation}
(D-2)I\!{\cal B}_{11}+\Omega^2I_{11}=(D-2)I\!{\cal
  B}_{22}+\Omega^2I_{22}=
(D-2)I\!{\cal
B}_{33}=\dots=(D-2)I\!{\cal B}_{DD}\label{vir2d}\,.
\end{equation}
The simplest solution of Eqs. (\ref{vir2d}) is the one with axial symmetry:
\begin{equation}
I_{11}=I_{22}\,,~~~~~I\!{\cal B}_{11}=I\!{\cal B}_{22}\,,,
\end{equation}
which is the generalization of MacLaurin sequence; the rotation
frequency is given by
\begin{equation} \Omega^2=(D-2)\frac{I\!{\cal B}_{33}-I\!{\cal B}_{11}}{I_{11}}\,, \end{equation}
and can be easily computed. But equations (\ref{vir2d}) leave open the possibility to have other, non
axisymmetric equilibrium configurations: the generalization of the Jacobi sequence. Their detailed study will
have to wait for other, possible numerical, studies.
\section{\label{sec:conclusions}Concluding remarks}

\subsection{\label{subsec:conclusions} Equilibrium configurations in higher dimensions}

We have started the investigation of equilibrium figures of rotating fluids in higher dimensions. We used two
appealingly simple models: one in which the fluid is held together by surface tension and the other in which
self-gravity is the cohesive force. We have developed the basic tools to start an investigation of this
important subject. Nevertheless, this work represents only a first step towards the understanding of the
behavior of rotating fluids in higher dimensions. Many questions remain to be addressed. It is clear that the
general behavior observed for $D=3$ still holds for higher $D$. It is natural to expect that (for instance for
fluid drops), a non axisymmetric figure bifurcates from the point where the axisymmetric figure starts to be
unstable, the point at which relation (\ref{instcond}) saturates. What is the shape of this figure? Is the
instability point really a bifurcation point? This would allow one to generalize Figure \ref{fig:twolobes} to
higher dimensions. Furthermore, our study must be extended to encompass more general perturbations and more
general angular momentum configurations. Another pressing issue is to determine whether there is always a
toroidal figure of equilibrium in general $D$ dimensions: could it be that for certain $D$ a drop {\it always}
encloses the origin regardless of its rotation? If so, this could be an excellent model for ultra-rotating
Myers-Perry black holes, for which the angular momentum is unbounded. Concerning self-gravitating spheroids,
there are many more unanswered questions. The formalism to study self-gravitating fluid drops must be further
developed to find the actual equilibrium shapes as well as their stability. This would lead to a generalization
of the MacLaurin and Jacobi sequences to higher dimensions. It would also be exciting to study numerically
self-gravitating toroidal configurations (the analogue of the Dyson rings).

\subsection{\label{subsec:fluidsblackholes} Fluids and black objects}

\subsubsection{\label{subsubsec:mimic} Mimicking event horizons with cohesive forces}

The work presented here also allows one to better understand how gravity (even in the strong field regime)
behaves in higher dimensions. Given the number of new solutions and instabilities discovered in this setting
over the last few years, this is a particularly fascinating topic. A lesson to learn from the examples shown
here (the Rayleigh-Plateau instability, the toroidal configurations, the DCF instability, the Dyson rings,
etecetera) is that many attractive forces, like Newtonian gravity and surface tension, seem to mimic strong
gravity situations very well. There is probably something more to it, when one shows that surface tension
mimics {\it so} well the behavior of horizons, but this remains to be explained. A particularly promising
proposal \cite{emparan}, which gathers support from our work, is that the physics of fluids and black holes can
be related semi-quantitatively for smooth enough objects. If the object is very inhomogeneous or is highly
distorted, then the gravitational self-interaction between different parts of the object will be strong, and
then one expects that the details of the (fully non-linear) dynamics of the interaction will matter. So, for
very inhomogeneous objects, general relativity or fluid dynamics would yield very different results. This means
that the fluid approach won't, in principle, be useful for cases such as: the deeply non-linear evolution of
the GL instability, when large inhomogeneities develop; the Myers-Perry black holes close to
the extremal singularity (in the cases they admit a Kerr bound); thick black rings, and in general black rings in the region where non-uniqueness
occurs. This is unfortunate since these are very intriguing and important regimes. Nevertheless, a plethora of
important and smooth objects (like the ones studied here) should then be described by these very simple models,
and could even be a starting point for more complex situations.

\subsubsection{\label{subsec:bifdiagram} The bifurcation diagram of black objects}

In the Introduction, the suggestion is made that bifurcation diagram in Fig. \ref{fig:brown} is in some sense
universal, the qualitative behavior holding for most objects held together by some cohesive force. The natural
generalization is to conjecture it may describe bifurcation diagrams of black objects.
 We would like to comment on the limitation of that generalization \footnote{We thank an
anonymous referee for suggesting this discussion.}.

\vskip 5mm 
{\it \large Kerr bound}\vskip 5mm

The first point to address concerns the existence of a Kerr bound. If the bifurcation diagram Fig.
\ref{fig:brown} were indeed universal, the angular momentum of black holes would be bounded. In fact, the
results in Section \ref{sec:eqshapesurface} imply that there is a bound on $J$ in {\it any} of the cases I and
II. We know however that the angular momentum of black holes in ``case I'' is unbounded (for $D>4$)
\cite{myersperry,emparanmyers}, whereas in case II it is bounded \cite{myersperry} (see for instance Eq. 7 in
\cite{Kunduri:2006qa}). Thus the surface tension model does not accurately describe ultra-spinning black holes,
for the cases studied numerically ($D=3,4,5$). Accurate numerical results are needed for higher dimensions, as
well as results for self-gravitating fluids. This would clear the issue.

\vskip 5mm 
{\it \large Instabilities}\vskip 5mm
The second point concerns the instability of the solutions. The results available in the literature concerning
black hole instability can be summarized as

(i) For $D>4$, case I, there should be both axisymmetric and non-axisymmetric instabilities
\cite{emparanmyers}.

(ii) For $D=4$ spatial dimensions, case I, the analysis of \cite{emparanmyers} does not apply; it was been
suggested in \cite{Emparan:2001wn} that there could be an axisymmetric instability before the Kerr bound,
ending in a black ring.

(iii) For case II, the study in \cite{Kunduri:2006qa} suggests no instability.

The studies in \cite{emparanmyers} and \cite{Kunduri:2006qa} focused only on a special kind of perturbation,
related to a traceless tensor mode. This seems to belong to the same representation of our ``toroidal mode'',
even if one relates to the metric and the other to the fluid.

In Section \ref{sec:stabilitysurface} we found that a ``case I'' rotating drop is unstable. This agrees with
the black hole picture (i) above. We were not able to extend the discussion to rotating drops in case II. So
far both pictures (black hole and fluids) agree. The long-known discrepancy is the $D=3$ case: Kerr black holes
in $D=3$ spatial dimensions are stable, whereas the fluid picture would predict instability. Another point of
discrepancy is that the most general vacuum solution of Einstein's equations in $3$ spatial dimensions is the
Kerr family, with spherical topology, so black rings are absent in this case. We can thus argue that $D=3$ is a
remarkable exception.

\vskip 5mm 
{\it \large Non-axisymmetric figures of equilibrium}\vskip 5mm
The final point we would like to touch upon is the question of existence of non-axisymmetric figures of
equilibrium. In the simple fluid models (see Figs. \ref{fig:brown} and \ref{fig:bifgrav}) there is a family of
non-axisymmetric stationary figures; one of the main points of our paper was to show that this holds for fluids
in general higher dimensions, too. We went on to find non-axisymmetric instability which we argued brings the
drop to a non-axisymmetric stationary configuration (in the case of gravitating fluid, we proved that there
exist stationary, non-axisymmetric configurations analogous to the Jacobi sequence). However, it is shown in
\cite{Hollands:2006rj} that higher dimensional rotating black objects must be axisymmetric. Therefore
stationary non-axisymmetric black solutions do not exist, and one would be led to think that the bifurcation
diagram for fluids is completely misleading.

The explanation for the disagreement is very simple, and lies in the fact that one is expecting from the model
something it was not built to do: a more realistic analogy would have to take emission of gravitational waves
into account, never considered up to now. Since non-axisymmetric objects radiate gravitational waves, we would
not expect to find stationary equilibrium solutions of that kind. Indeed, for three spatial dimensions this
problem was considered by Chandrasekhar \cite{chandragw}. Chandrasekhar proved that when one takes emission of
gravitational waves into account (all other ``curved spacetime'' effects being neglected), the Jacobi
configuration is unstable and evolves in the direction of increasing angular velocity, approaching the
bifurcation point. Furthermore, he proved that radiation reaction does not make the fluid unstable past the
point of bifurcation. Thus the bifurcation diagram for black objects would be drawn by first deleting all the
non-axisymmetric figures.

It is reasonable to think that this result, i.e. the non-existence of
non-axisymmetric stationary fluid configurations in general
relativity, can be generalized to higher dimensions. If this is the
case, this would be the analogue of the no-go result of
\cite{Hollands:2006rj} for higher dimensional rotating black holes.

\subsection{\label{subsec:worktodo} Future work}
The two examples we have studied (surface tension and self-gravitating fluids) can therefore improve and
strengthen our understanding of black objects. The fact that rotating fluid drops develop instabilities at
finite (usually small) rotation seems to suggest that black holes cannot have large rotation. This was also put
forward by Emparan and Myers \cite{emparanmyers} in connection with the Gregory-Laflamme instability. The
examples shown here and the classical three-dimensional results discussed in the Introduction suggest that the
final state of this instability is a non axi-symmetric figure. Some evidence was put forward already in
\cite{emparanmyers} and this work adds some more evidence to that possibility. In fact, it might not even be
possible to find matter configurations with high enough rotation to form these ultra-spinning black holes
following collapse: we prove in Appendix \ref{app:poincareinequality} that self-gravitating, rotating
homogeneous fluids always have a bounded angular velocity (this is a generalization of a classic result by
Poincar\'e). Future work along these lines (mimicking strong gravity situations with analogue models) could
start by investigating the black string phase transition \cite{mu-inpark,Gubser:2001ac,Sorkin:2004qq} within
either the surface tension or the self-gravity models. Another potentially interesting topic would be to
investigate the collapse of toroidal fluid configurations to form black rings.
\vskip 2cm
\section*{Acknowledgements}
We warmly thank Emanuele Berti, Marco Cavagli\`a, \'Oscar Dias, Jos\'e Lemos and Mu-In
Park for a careful and critical reading of the manuscript and for many useful comments.
We are indebted to Roberto Emparan for very useful suggestions and specially for having
enlightened us as to the significance of some of the results. We are also indebted to
Dr. Claus-Justus Heine for very useful correspondence on this subject and for preparing
Figure \ref{fig:twolobes} for us. We thank Dr. Robert Brown for kind permission to use
Figure \ref{fig:brown} and Drs. Yoshiharu Eriguchi and Izumi Hachisu for kind
permission to use Figure \ref{fig:bifgrav}. VC acknowledges financial support from Funda\c c\~ao
Calouste Gulbenkian through the Programa Gulbenkian de Est\'{\i}mulo \`a Investiga\c
c\~ao Cient\'{\i}fica.
\appendix

\section{\label{app:surfacetensiondrop} Oscillation modes of a spherical drop held together by surface tension}
What are the oscillation modes of a spherical drop held together by
surface tension? In three spatial dimensions the answer to this
classical problem can be found for instance in \cite{lamb,reid}. We
will now generalize it to an arbitrary number of dimensions. Take a
spherical drop with radius $R_0$. Define again
\begin{equation} \Upsilon \equiv \frac{\delta P}{\rho}\,,\label{upsilonsurface2} \end{equation}
where $\delta P$ is the change in pressure inside the drop (the
outside pressure is taken to be zero), after it is perturbed by a
displacement of the form
\begin{equation} r=R_0+\epsilon e^{\omega t}Y({\rm angles})\,, \label{pert2surface} \end{equation}
where $Y({\rm angles})$ are hyper-spherical harmonics, with eigenvalue
$l(l+D-2)$. The equations governing small perturbations are
\begin{eqnarray} P_0+\delta P&=&T{\rm div}\vec{n}\label{46e} \,,\\ \frac{\partial
\bf u}{\partial t}&=&-{\rm grad} \Upsilon \label{gradups2surface}\,,\\
{\rm div} {\bf u}&=&0\,,\label{divergenless2surface}\\ \Upsilon
&\equiv& \frac{\delta P}{\rho} \label{defups2surface}\,. \end{eqnarray}
With the displacement (\ref{pert2surface}), equation (\ref{46e}) can
be solved to yield (for details we refer the reader to Landau and
Lifshitz \cite{reid}) near $r=R_0$ and to first order in $\epsilon$
\begin{equation} \frac{\delta P}{\rho}=-\frac{T}{R_0^2\rho}(l-1)(l+D-1) \epsilon
e^{\omega t}Y({\rm angles})\,.\label{deltaP2surface} \end{equation}
Therefore
\begin{equation} \Upsilon =-\frac{T (l-1)(l+D-1)}{\rho R_0^{l+2}}r^l \epsilon
e^{\omega t}Y({\rm angles})\,.\label{aiai}\end{equation}
On the other hand, observing that $\nabla^2\left (\frac{\delta
P}{\rho}\right )=0$ we may immediately write
\begin{equation} \frac{\delta P}{\rho}=\epsilon F_p\frac{r^l}{R_0^l} Y({\rm
  angles})e^{\omega t}\,, \end{equation}
with $F_p$ a constant to be determined. We thus have from the
definition of $\Upsilon$ that
\begin{equation} \Upsilon =\frac{r^l}{R_0^l}\epsilon \, F_p\,Y({\rm
angles})e^{\omega t}\,.\label{upsxxx}\end{equation}
Equation (\ref{gradups2surface}) then yields
\begin{equation} u_r=-\frac{1}{\omega}\frac{d}{dr}\Upsilon\,,\label{ur2surface} \end{equation}
Now, for this velocity to be compatible with the radial perturbation
(\ref{pert2surface}) one must have
\begin{equation} F_p=-\frac{R_0\omega^2}{l} \,. \label{epsilon2surface} \end{equation}
Inserting this in (\ref{upsxxx}) and equating the resulting expression
with (\ref{aiai}) we get
\begin{equation} \omega^2=-\frac{T l(l-1)(l+D-1)}{\rho R_0^3} \,, \end{equation}
which is a generalization of the three-dimensional result found in \cite{lamb,reid}. Note that the drop is
always stable ($\omega$ is imaginary) and that $l=0$ and $1$ modes do not exist. The first corresponds to
spherically symmetric pulsation of the drop, which violates the constant volume assumption and the second
corresponds to a translation of the drop as whole.
\section{\label{app:gravitysphere} Oscillation modes of a self-gravitating,
non-rotating sphere}
Here we analyze the modes of a self-gravitating fluid sphere, in which
case we can do without the virial formalism and find an exact
analytical expression for the oscillation frequencies. We aim at
generalizing the four-dimensional result by Kelvin and others
\cite{lamb}. The analysis from Section \ref{sec:cylindergravity}
carries over with minimal changes.

The gravitational potential ${\cal B}({\bf x})$ inside and outside the
globe satisfy respectively
\begin{eqnarray}
\nabla^2{\cal B}({\bf x})^i&=&-(D-2)C_D G \rho\,,\label{visphere}\\
\nabla^2{\cal B}({\bf x})^e&=&0\,. \label{vesphere} \end{eqnarray}
In the unperturbed state we get
\begin{eqnarray}
{\cal B}({\bf x})^i_0&=& -\frac{(D-2)C_D G\rho}{2D}\left (-
\frac{DR_0^2}{D-2}+r^2\right )\,,\label{Vspherei}\\
{\cal B}({\bf x})^e_0&=&\frac{C_D G\rho R_0^D}{D}
\frac{1}{r^{D-2}}\,.
\end{eqnarray}
Hydrostatic equilibrium demands that the pressure $P_0$ satisfies
\begin{equation}
\frac{d}{dr}\left (\frac{P_0}{\rho}-{\cal B}({\bf x})\right )=0
\,,
\end{equation}
which yields, upon requiring $P_0(r=R_0)=0$,
\begin{equation} P_0=\frac{(D-2)C_D G \rho ^2}{2D}(R_0^2-r^2)\,. \end{equation}
Suppose now the system is subjected to a perturbation, of general form
\begin{equation}
r=R_0+\epsilon e^{\omega t}Y({\rm angles})\,, \label{pert2}
\end{equation}
where $Y({\rm angles})$ are hyper-spherical harmonics, with eigenvalue
$l(l+D-2)$.

The equations governing small perturbations are
\begin{eqnarray} \nabla^2 \delta {\cal B}({\bf x})&=&0\,,\\ \frac{\partial \bf
u}{\partial t}&=&-{\rm grad} \Upsilon \label{gradups2}\,,\\ {\rm div}
{\bf u}&=&0\,,\label{divergenless2}\\ \Upsilon &\equiv& \frac{\delta
P}{\rho}-\delta {\cal B}({\bf x})^i
\label{defups2}\,. \end{eqnarray}
Here ${\bf u}$ stands for the velocity of the perturbed fluid. Let's
solve for the first of these equations. We find
\begin{eqnarray}
{\cal B}({\bf x})^e&=&{\cal B}({\bf x})^e_0+\epsilon  A r^{-l-D+2}
Y({\rm angles})e^{\omega t} \,,\\
{\cal B}({\bf x})^i&=&{\cal B}({\bf x})^i_0+\epsilon  Br^{l}
Y({\rm angles})e^{\omega t}\,. \end{eqnarray}
The constants $A$ and $B$ are determined by requiring continuity of
the potential and its derivative on the boundary. By keeping only
dominant terms in $\epsilon$ we find
\begin{eqnarray}
A&=&\frac{(D-2)C_D G \rho}{2l+D-2}R_0^{l+D-1}\,,\\
B&=&\frac{(D-2)C_D G \rho}{2l+D-2}R_0^{1-l}\,. \end{eqnarray}
Therefore we have
\begin{equation} \delta {\cal B}({\bf x})^i=\epsilon \frac{(D-2)C_D G \rho}{2l+
D-2}R_0^{1-l} r^{l}Y({\rm angles})e^{\omega t}\,. \end{equation}
Observing that $\nabla^2\left (\frac{\delta P}{\rho}\right )=0$ we may
immediately write
\begin{equation} \frac{\delta P}{\rho}=\epsilon F_p\frac{r^l}{R_0^l} Y({\rm
  angles})e^{\omega t}\,, \end{equation}
with $F_p$ a constant to be determined. We thus have from the
definition of $\Upsilon$, equation (\ref{defups2}), that
\begin{equation} \Upsilon \frac{r^l}{R_0^l}\epsilon \left (F_p-\frac{(D-2)C_D G
\rho R_0}{2l+D-2}\right )Y({\rm angles})e^{\omega t}\,.\end{equation}
Equation (\ref{gradups2}) then yields

\begin{equation} u_r=-\frac{1}{\omega}\frac{d}{dr}\Upsilon\,,\label{ur2} \end{equation}
Now, for this velocity to be compatible with the radial perturbation
(\ref{pert2}) one must have
\begin{equation} F_p=\frac{(D-2)C_D G \rho R_0}{2l+D-2}-\frac{R_0\omega^2}{l}
\,. \label{epsilon2} \end{equation}
Imposing a vanishing pressure on the surface of the globe we have finally
\begin{equation} \omega^2=-\frac{2(D-2)C_D G \rho}{D}\frac{l(l-1)}{2l+D-2}\,,
\label{oscglobe} \end{equation}
which is a generalization of Kelvin's four-dimensional result \cite{lamb} to an arbitrary number of dimensions.
As in the previous Section, the globe is always stable and $l=0$ and $l=1$ modes do not exist. Result
(\ref{oscglobe}) can be compared with the results for the modes of spherically symmetric black holes in higher
dimensions \cite{modeshigherdim,modeshigherdim2}.

\section{\label{app:poincareinequality} An upper bound for the rotation
of an homogeneous fluid mass}
In four dimensions there is an upper bound, derived by Poincar\'e \cite{poincaresurfacetension} for the maximum
angular velocity of a rigidly rotating, homogeneous fluid mass in equilibrium. Consider a region $V$ in three
dimensional space bounded by a surface $S$. Let $\frac{\partial}{\partial n_e}$ denote differentiation of a
function across $S$ in the direction of the unit normal $n_e$ outward from $V$. The divergence theorem states
that for any (sufficiently smooth) vector field $\vec{W}$
\begin{equation} \int_S\vec{W}.\vec{n}\,d\sigma=\int_V {\rm div}\vec{W}d^3x\,. \end{equation}
If we choose $\vec{W}={\rm grad}{\cal B}$, where ${\cal B}$ is the
gravitational potential, we get
\begin{equation} \int_S\frac{\partial {\cal B}}{\partial n_e}\,d\sigma=\int_V \nabla^2{\cal B}d^3x=-\int_V4\pi G \rho
d^3x\,, \end{equation}
where we used Poisson's equation in the last equality.

A necessary (but not sufficient) condition for equilibrium is that the
force at every point in $S$ be directed inward. But the total force is
$F_r=\frac{\partial U}{\partial n_e}$, where $U$ is the modified potential
\begin{equation}
{\cal U}={\cal B}+\frac{1}{2}\Omega^2\left ((x^1)^2+(x^2)^2\right )
\end{equation}
(we choose $x^3=z$ to be the axis of rotation). We then must require
\begin{equation} \frac{\partial{\cal U}}{\partial n_e}<0\,\,\Rightarrow \int_S \frac{\partial {\cal U}}{\partial
n_e}d\sigma=\int_V \nabla^2 U d^3x =\int_V\left(-4\pi G\rho+2\Omega^2\right)d^3x <0\,. \end{equation}
thus we have the Poincar\'e's inequality \footnote{A more
sophisticated analysis by Crudelli \cite{crudelli} sharpened this
inequality to half its value.}
\begin{equation}
\Omega^2<2\pi \rho G\,.
\end{equation}
This derivation can be easily generalized to $D>3$ dimensional
Euclidean space. If we assume the gravitational potential is
(\ref{gravd})
\begin{equation} {\cal B}({\bf x})=G\int d^Dx'\frac{\rho({\bf x'})}{|{\bf x}-
{\bf x'}|^{D-2}}\,,
\end{equation}
then Poisson's equation becomes
\begin{equation} \nabla^2{\cal B}=-(D-2)C_DG\rho \,,\label{glaplacian}\end{equation}
where
\begin{equation} C_D=\frac{2\pi^{D/2}}{\Gamma\left(\frac{D}{2}\right)} \,,
\label{areasphere}
\end{equation}
and $C_Dr^{D-1}$ is the area of the $(D-1)$-sphere. Indeed,
\begin{equation} \int_V\nabla^2\left(\frac{1}{r^{D-2}}\right)d^Dx= \int_S
\frac{\partial}{\partial
n_e}\left(\frac{1}{r^{D-2}}\right)d\sigma =-(D-2)\int_S\frac{x^i}{r^D}
\frac{x^i}{r}d\sigma=-(D-2)C_D\,. \end{equation}
We now consider the two cases labeled by I and II, that is,
\begin{itemize}
\item[I:] one non-vanishing angular momentum;
\item[II:] all angular momenta coincident, and $D$ odd.
\end{itemize}

The modified potential, in both cases, writes as
\begin{equation} {\cal U}={\cal B}+\frac{1}{2}\Omega^2\sum_a(x^a)^2 \,,\end{equation}
where $a=1,2$ in case I, $a=1,\dots,D-1$ in case II. Its Laplacian is
\begin{eqnarray}
\nabla^2 {\cal U}&=&-C_D\rho G+2\Omega^2~~~~~~~~~~~~\,\hbox{in case I}\nonumber\,,\\
\nabla^2 {\cal U}&=&-C_D\rho G+(D-1)\Omega^2~~~~~\hbox{in case II}\,, \end{eqnarray}
therefore Poincar\'e's inequality becomes
\begin{eqnarray}
\Omega^2&<&\frac{C_D}{2}\rho G~~~~~\hbox{in case I}\nonumber\,,\\
\Omega^2&<&\frac{C_D}{D-1}\rho G~~~~~\hbox{in case II}\,. \end{eqnarray}
For angular velocities larger than this threshold, the gravitational pull is insufficient to balance the
centrifugal force, and there body is disrupted.
\section{Toroidal perturbations of a rotating drop}\label{toroidal}
Here we compute some useful quantities in the case of a rotating drop
with one non-vanishing angular momentum, constant density $\rho$ and
with toroidal perturbations (following the denomination of
Chandrasekhar \cite{chandra}), i.e.
\begin{eqnarray}
\xi^a&=&T^{ab}x^b\nonumber\,,\\
\xi^A&=&0\nonumber\,,\\
T^{ab}&=&\left(\begin{array}{cc}\alpha&\beta\\\beta&-\alpha\,,\\
\end{array}\right)\,,\label{tor1}
\end{eqnarray}
with $\alpha,\beta$ constant parameters. We remind that, in the case
with one non-vanishing angular momentum, the coordinates $\{x^i\}$ are
decomposed as in (\ref{coordinatesa}), (\ref{deftoz1})
\begin{equation}
x^i=(x^a,x^A)~;~~~~a=1,2~;~~A=3,\dots,D
\end{equation}
and we have defined
\begin{eqnarray}
\varpi&=&\sqrt{(x^1)^2+(x^2)^2}\nonumber\,,\\
z&=&\sqrt{(x^3)^2+\dots+(x^D)^2}\,,
\end{eqnarray}
so that
\begin{eqnarray}
x^1&=&\varpi\cos\theta\nonumber\,,\\
x^2&=&\varpi\sin\theta\,.
\end{eqnarray}
The $D$-dimensional volume integration element is
\begin{equation}
d^Dx=\varpi z^{D-3}d\varpi d\theta dzd\Omega_{D-3}\,.
\end{equation}
If we integrate on the angular variables, we get
\begin{equation}
\int_{S^1\times S^{D-3}}d^Dx=2\pi C_{D-2} \varpi z^{D-3}d\varpi dz\,,
\end{equation}
where $C_{D-2}$ is the area of a unit $(D-3)$-sphere (\ref{areasphere}).

Let us now consider the $(D-1)$-dimensional surface $S$ of equation $z= f(\varpi)$. Since on $S$, $dz=\phi
d\varpi$, the metric element on $S$ is
\begin{equation}
d\sigma^2=(1+\phi)^2d\varpi^2+\varpi^2d\theta^2+
[f(\varpi)]^2d\Omega_{D-3}^2\,.
\end{equation}
The surface integration element is then
\begin{equation}
dS=\sqrt{1+\phi^2}\varpi [f(\varpi)]^{D-3}
d\varpi d\theta d\Omega_{D-3}\,.
\end{equation}
In particular, being
\begin{equation}
n^a=-\frac{\phi x^a}{\sqrt{1+\phi^2}\varpi}\,,
\end{equation}
we have
\begin{equation}
dS^a=n^adS=-\phi x^a[f(\varpi)]^{D-3}d\varpi d\theta d\Omega_{D-3}\,.
\end{equation}
After integration on the $(D-3)$-sphere we get,
\begin{equation}
\int_{S^{D-3}}dS^a=-\phi x^a[f(\varpi)]^{D-3}C_{D-2}d\varpi d\theta\,.
\label{dSa}
\end{equation}
Notice that in the $D=3$ case we have $dS^a=-2\phi x^ad\varpi d\theta$ where the factor $2$ accounts for the
two hemispheres of the drop.

\subsection{Virial integrals}
We can easily compute the virial integrals
\begin{equation}
\Gamma^{i;j}=\int_V\rho\xi^ix^jd^Dx\,.
\end{equation}
From (\ref{tor1}) it follows $\Gamma^{A;j}=0$; furthermore,
\begin{equation}
\int_V x^ix^jd^Dx=0~~~~{\rm if}~i\neq j
\end{equation}
because the drop is symmetric with respect to reflections to any axis,
therefore $\Gamma^{a,A}=0$.

We have
\begin{eqnarray}
\int_V(x^1)^2d^Dx&=&\left(\int_0^{2\pi}d\theta\cos^2\theta\right) C_{D-2}~
\int_0^ad\varpi\varpi^3~\int_0^{f(\varpi)}
dzz^{D-3}\nonumber\,,\\
&=&\pi\frac{C_{D-2}}{D-2}\int_0^ad\varpi\varpi^3 \left[f(\varpi)\right]^{D-2}=\int_V(x^2)^2d^Dx\,,
\end{eqnarray}
where $\varpi=a$ is the intersection of the drop with the $x^1-x^2$ plane
(namely, $f(a)=0$); therefore,
\begin{eqnarray}
\Gamma^{1;1}&=&-\Gamma^{2;2}=\pi\rho\alpha\frac{C_{D-2}}{D-2}
\int_0^a\varpi^3\left[f(\varpi)\right]^{D-2}d\varpi\,,\\
\Gamma^{1;2}&=&\Gamma^{2;1}=\pi\rho\beta\frac{C_{D-2}}{D-2}
\int_0^a\varpi^3\left[f(\varpi)\right]^{D-2}d\varpi\,.
\end{eqnarray}
We notice that these formulae are valid for every $D\ge 3$. When
$D>3$, $C_{D-2}$ represents the area of the unit $(D-3)$-sphere;
when $D=3$, the factor $C_1=2$ accounts for the two hemispheres
composing the drop.

\subsection{Surface energy integrals}
Here we want to compute the quantities $\delta\mathfrak S^{12}$ and
$\delta\mathfrak S^{11}-\delta\mathfrak S^{22}$ where (see (\ref{eqkS}))
\begin{equation}
\delta\mathfrak S^{ab}=-T\left[\int_S\xi^bn^k_{~,k}dS^a+\int_Sx^b \delta(n^k_{~,k})dS^a
-\int_Sx^bn^k_{~,k}\frac{\partial \xi^c}{\partial x^a}dS^c\right]\,,\label{eqkS1}
\end{equation}
and
\begin{equation}
dS^a=-\phi x^a[f(\varpi)]^{D-3}C_{D-2}d\varpi d\theta\,.
\end{equation}
Being $\xi^a=T^{ab}x^b$, we have that
\begin{eqnarray}
&&\int d\theta\left[\xi^2x^1-x^2\frac{\partial\xi^c}{\partial x^1}x^c
\right]=\int_0^{2\pi} d\theta[T^{2a}x^ax^1-T^{1a}x^2x^a]=
\int_0^{2\pi} d\theta\left((x^1)^2-(x^2)^2\right)\beta=0\nonumber\,,\\
&&\nonumber\\
&&\int_0^{2\pi} d\theta\left[\xi^1x^1-\xi^2x^2-x^1\frac{\partial\xi^c}{\partial x^1}x^c +x^2\frac{\partial\xi^c}{\partial
x^2}x^c\right] =\int_0^{2\pi} d\theta\left[T^{1a}x^ax^1-T^{2a}x^ax^2
-x^1T^{c1}x^c+x^2T^{c2}x^c\right]=0\nonumber\,.
\end{eqnarray}
therefore the first and third integrals in (\ref{eqkS1})
cancel for the components $\delta\mathfrak S^{12}$ and
$\delta\mathfrak S^{11}-\delta\mathfrak S^{22}$:
\begin{eqnarray}
\delta\mathfrak S^{12}&=&-T\int_S\delta(n^k_{~,k})x^2dS^1\nonumber\,,\\
\delta\mathfrak S^{11}-\delta\mathfrak S^{22}
&=&-T\int_S\delta(n^k_{~,k})x^1dS^1-T\int_S\delta(n^k_{~,k})x^2dS^2\,.
\end{eqnarray}
The perturbation of the divergence $\delta(n^i_{~,i})$ is given
by (\ref{expdiv})
\begin{eqnarray}
\delta(n^i_{~,i})&=&(\delta n^i)_{,i}-\frac{\partial n^i}{\partial x^k}
\frac{\partial \xi^k}{\partial x^i}=(\delta n^a)_{,a}-T^{ab}n^a_{~,b}
+(\delta n^A)_{,A}\nonumber\,,\\
\delta n^a&=&\left(n^kn^l\frac{\partial \xi^k}{\partial x^l}\right)n^a
-n^j\frac{\partial \xi^j}{\partial x^a}=[n^bn^cT^{bc}]n^a-T^{ab}n^b\nonumber\,,\\
\delta n^A&=&\left(n^kn^l\frac{\partial \xi^k}{\partial x^l}\right)n^A =[n^bn^cT^{bc}]n^A\,.
\end{eqnarray}
with
\begin{equation}
n^a=-\phi\frac{x^a}{\varpi\sqrt{1+\phi^2}}\,,~~~~~n^A=
\frac{x^A}{z\sqrt{1+\phi^2}}\,.
\end{equation}
We have
\begin{equation}
n^a_{~,b}=-\frac{\phi'}{(1+\phi^2)^{3/2}}\frac{x^ax^b}{\varpi^2}
+\frac{\phi}{\sqrt{1+\phi^2}}\frac{x^ax^b}{\varpi^3}-\delta^{ab} \frac{\phi}{\sqrt{1+\phi^2}\varpi}\,,
\end{equation}
then we find
\begin{equation}
\delta\left (n^i_{~,i}\right )=-\frac{x^ax^bT^{ab}}{\varpi^2}
\left[\frac{\phi'(\phi^2-2)}{(1+\phi^2)^{5/2}}+\phi\frac{2
+3\phi^2}{\varpi(1+\phi^2)^{3/2}}-\frac{D-3}{z}
\frac{\phi^2}{(1+\phi^2)^{3/2}} \right ]\,.
\end{equation}
Therefore,
\begin{equation}
\delta\left (n^i_{~,i}\right )=-\left [\alpha \left ((x^1)^2-(x^2)^2 \right )+2\beta x^1 x^2\right ]Q(\varpi)
=-\varpi^2\left(\alpha\cos 2\theta+\beta\sin 2\theta\right) Q(\varpi)\,, \label{deltadiv1}
\end{equation}
where
\begin{equation}
Q=\frac{1}{\varpi ^2}\left
(\frac{\phi'(\phi^2-2)}{(1+\phi^2)^{5/2}}+\phi\frac{2
+3\phi^2}{\varpi(1+\phi^2)^{3/2}}-\frac{D-3}{z}
\frac{\phi^2}{(1+\phi^2)^{3/2}} \right )\,.
\label{QQ1}
\end{equation}
The surface integrals $\delta\mathfrak S^{12}$, $\delta\mathfrak
S^{11}-\delta\mathfrak S^{22}$ are then
\begin{eqnarray}
\delta\mathfrak S^{12}&=&TC_{D-2}\int_0^a d\varpi\int_0^{2\pi}
d\theta\phi x^1x^2[f(\varpi)]^{D-3}\delta
(n^k_{~,k})\nonumber\,,\\
&=&-\frac{T}{2}\beta\pi C_{D-2}\int_0^a d\varpi
\phi\varpi^4Q(\varpi)[f(\varpi)]^{D-3}\,,\\
\delta\mathfrak S^{11}-\delta\mathfrak S^{22}
&=&TC_{D-2}\int_0^a d\varpi\int_0^{2\pi} d\theta
\phi\left[(x^1)^2-(x^2)^2\right][f(\varpi)]^{D-3}\delta(n^k_{~,k})
\nonumber\,,\\
&=&-T\alpha\pi C_{D-2}\int_0^a
d\varpi\phi\varpi^4Q(\varpi)[f(\varpi)]^{D-3}\,.
\end{eqnarray}
\section{\label{app:newton} Newton's constant}
Here we would like to relate our adopted definition of Newton's
constant to the one adopted in the by now classical work by Myers and
Perry \cite{myersperry}. They define a Newton's constant $G_{MP}$ such
that Einstein's equations read
\begin{equation} R_{\mu \nu}-\frac{1}{2}g_{\mu \nu}R=8\pi G_{MP}T_{\mu \nu}\,. \end{equation}
We will now derive ``Newton's law'' from this equation, in order to
relate the different Newton constants. To derive Newton's law
$a=-\nabla {\cal B}({\bf x})$, we consider the Newtonian limit of
Einstein's equation (slowly moving particles, weak and static
gravitational fields).

A test particle follows the geodesics of the spacetime
\begin{equation} \frac{d^2x^{\mu}}{d\tau ^2}+\Gamma^{\mu}_{\rho
\sigma}\frac{dx^{\rho}}{d\tau}\frac{dx^{\sigma}}{d\tau} =0\,.\end{equation}
Under the ``moving slowly'' condition $\frac{dx^i}{d\tau}\ll
\frac{dt}{d\tau}$ this is simplified to
\begin{equation} \frac{d^2x^{\mu}}{d\tau ^2}+\Gamma^{\mu}_{00}\left
(\frac{dt}{d\tau}\right )^2=0\,. \end{equation}
Due to the staticity assumption,
\begin{equation} \Gamma^{\mu}_{00}=\frac{1}{2}g^{\mu \lambda}\left
(\partial_0g_{\lambda 0}+\partial_0g_{0
\lambda}-\partial_{\lambda}g_{0 0}\right )=-\frac{1}{2}g^{\mu
\lambda}\partial_{\lambda}g_{0 0}\,, \end{equation}
so that the geodesic equation simplifies to
\begin{equation} \frac{d^2x^{\mu}}{d\tau ^2}-\frac{1}{2}g^{\mu
\lambda}\partial_{\lambda}g_{0 0}\left (\frac{dt}{d\tau}\right
)^2=0\,. \end{equation}

Lastly, the Newtonian limit implies weak gravitational fields, so we set
\begin{equation} g_{\mu \nu}=\eta_{\mu \nu}+h_{\mu \nu}\,,\quad |h_{\mu \nu}|\ll
1\,. \end{equation}
Thus we have (we can show that to first order $g^{\mu \nu}=\eta^{\mu
\nu}-h^{\mu \nu}$)
\begin{equation} \Gamma^{\mu}_{00}=-\frac{1}{2}\eta^{\mu
\lambda}\partial_{\lambda}h_{0 0}\,, \end{equation}
and therefore
\begin{equation} \frac{d^2x^{\mu}}{d\tau ^2}=\frac{1}{2}\eta^{\mu
\lambda}\partial_{\lambda}h_{0 0}\left (\frac{dt}{d\tau}\right
)^2\,.\end{equation}
Dividing the spatial components of this equation by $\left (\frac{dt}{d\tau}\right )^2$ we have finally
\begin{equation} \frac{d^2x^i}{dt^2}=\frac{1}{2}\partial_ih_{00} \,,\end{equation}
which allows one to identify $h_{00}=-2{\cal B}({\bf x})$. On the other hand, from Myers and Perry's equation
(1.5), we get
\begin{equation} \nabla^2 h_{00}=-16\pi\frac{D-2}{D-1}G_{MP}\label{gmp}\,. \end{equation}
Comparing our (\ref{glaplacian}) with (\ref{gmp}), we get the relation between our adopted $G$ and $G_{MP}$ as
(there's an extra minus sign that follows from the conventions adopted here for the relation force-potential)
\begin{equation} C_D G=\frac{8\pi}{D-1} G_{MP} \,.\end{equation}
%

\end{document}